\documentclass[aps,twocolumn]{revtex4}\voffset=2cm
\usepackage{graphicx}

\newcommand{\be}{\begin{equation}}
\newcommand{\ee}{\end{equation}}
\newcommand{\bea}{\begin{eqnarray}}
\newcommand{\eea}{\end{eqnarray}}

\begin{document}
\title{Winding angle distribution for planar random walk, polymer
ring entangled with an obstacle, and all that:
Spitzer-Edwards-Prager-Frisch model revisited}

\author{A. Grosberg$^{1,2}$ and H. Frisch$^{3,4}$}
\affiliation{$^1$Department of Physics, University of Minnesota,
Minneapolis, MN 55455, USA \\  $^2$Institute of Biochemical
Physics, Russian Academy of Sciences, Moscow 117977, Russia
\\ $^3$Department of Chemistry, State University of New York at Albany, Albany, NY 12222, USA \\ $^4$
Physics Department, Stellenbosch University, Matieland, 7602 South
Africa}
\date{\today}

\begin{abstract}
Using a general Green function formulation, we re-derive, both,
(i) Spitzer and his followers results for the winding angle
distribution of the planar Brownian motion, and (ii)
Edwards-Prager-Frisch results on the statistical mechanics of a
ring polymer entangled with a straight bar.  In the statistical
mechanics part, we consider both cases of quenched and annealed
topology.  Among new results, we compute exactly the (expectation
value of) the surface area of the locus of points such that each
of them has linking number $n$ with a given closed random walk
trajectory ($=$ ring polymer).  We also consider the
generalizations of the problem for the finite diameter (disc-like)
obstacle and winding within a cavity.  \\ PACS numbers: 61.41.+e,
36.20.Ey, 87.15.Cc
\end{abstract}

\maketitle

\section{Introduction}

\begin{figure}
\centerline{\scalebox{0.8} {\includegraphics{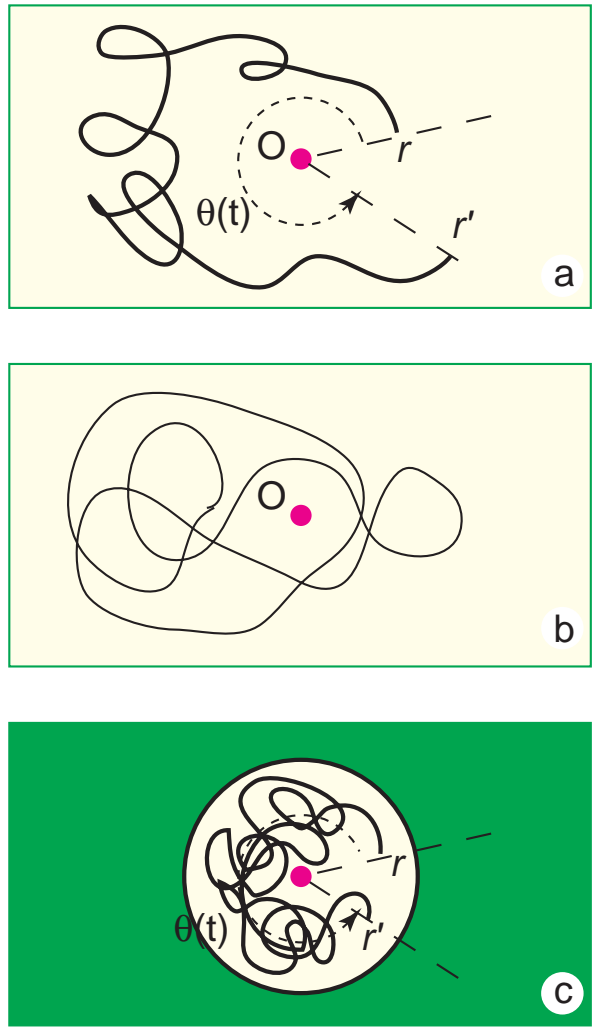}}}
\caption{Schematic representattion of the model.  (a) Random walk
winding around an obstacle ${\cal O}$.  This obstacle might be
just a point, or it might be a disc of a finite radius $b$.  (b)
Closed polymer winding around an obstacle.  Mathematically, this
is similar to (a), except both ends are kept together.  (c)
Similar to (a), except the trajectory cannot leave a ``cavity'' of
some radius $B$. }\label{fig:winding}
\end{figure}

In 1958, Spitzer \cite{Spitzer} established the following result.
Consider the two-dimensional random walk starting at a point other
than ${\cal O}$, and let $\theta (t)$ be the total continuous
angle wound by the walker around ${\cal O}$ up to time $t$ (see
figure \ref{fig:winding} a). The \emph{Spitzer law} says that the
quantity $ \theta (t) / \ln t $ at large enough $t$ is Lorenz (or
Cauchy) distributed:
\be W( \theta) = \frac{1}{\pi} \frac{1}{1+ x^2} \ ; \ x = \frac{2
\theta (t) }{ \ln t } \ . \label{eq:Spitzer} \ee
With a remarkable delay of about 25 years, a large group of
followers studied this law in depth
\cite{Yor,Privman,PitmanYor1,RudnickHu1,RudnickHu2,Duplantier_winding,PitmanYor2,Belisle,Comtet,Drossel,Samokhin1,Samokhin2}.
The central finding of these studies attributes the divergent
moments of the Spitzer distribution (\ref{eq:Spitzer}), e.g.
$\langle \theta^2 \rangle$, to the small scale properties of the
regular random walk trajectories.  Simply speaking, infinitely
large winding is accumulated while the trajectory is wandering
infinitely close to the obstacle ${\cal O}$.  Accordingly, this
pathology of divergent moments is removed by incorporating any
kind of ``granularity,'' or short length scale cut-off, in the
model.  Such modification of the model can be achieved in quite a
few ways. One way is to consider the random walk on the lattice
instead of the continuous space \cite{Belisle}; another way is to
look at the winding around a finite obstacle, say, a disc of some
radius $b$ \cite{Yor,PitmanYor1,RudnickHu1}; one more possibility
is to examine a broken line of straight segments of finite length
$b$ each instead of standard Wiener-measured random walk; yet
another way is to consider a worm-like smooth curve with an
effective segment $b$ (that is, the curve which adopts smoothly
curved shapes ${\bf r}(s)$ with the weight proportional to $\exp
\left[ - (b /2) \int \ddot{{\bf r}} ^2 ds \right]$, where $s$ is
the arc length). In all of these cases, winding is characterized
by the non-pathological distribution
%
%
%
\be W( \theta ) = \frac{\pi}{4 \cosh^2 ( \pi x / 2) }  \ ; \ x =
\frac{2 \theta (t) }{ \ln t } \ . \label{eq:NO_Spitzer} \ee
A similar distribution is also characteristic for the winding of
the self-avoiding walk
\cite{Privman,RudnickHu2,Duplantier_winding}; self-avoidance, in
this case, is just another way to suppress infinite winding at
infinitely small length scale. Mathematically, it turns out that
the winding angle distribution is in fact an example of a broad
class of limiting laws for the two-dimensional random walk
\cite{PitmanYor1,PitmanYor2}.

Many studies of winding angle distribution
\cite{Yor,Privman,RudnickHu1,RudnickHu2,Duplantier_winding,Comtet,Samokhin1,Samokhin2}
claim that entanglement of long polymer filaments is (one of)
their motivation(s).  Indeed, the relation to polymer physics does
exist.  It was found in 1967, almost a decade after Spitzer
\cite{Spitzer}, by Edwards \cite{Edwards} and, independently, by
Prager and Frisch \cite{Prager_Frisch} (see also an influential
review \cite{Wiegel}).  These authors came up with the model of a
polymer chain wound around a straight bar and topologically
entangled with this bar.  Given the analogy of a polymer chain
conformation with the random walk trajectory, the
Edwards-Prager-Frisch model is essentially the same as that
examined by Spitzer \cite{Spitzer}.  Neither of the works
\cite{Edwards,Prager_Frisch,Wiegel} makes a reference to
\cite{Spitzer}.  Most likely, mathematical work \cite{Spitzer} was
not known to physicists at the time, but even apart from that,
authors of the works \cite{Edwards,Prager_Frisch,Wiegel} did not
examine winding angle distribution for the random walk with open
ends, their goal was obviously to compute quantities similar to
those of physical interest for real polymers.  Unfortunately, no
explicit formula was obtained in the works
\cite{Edwards,Prager_Frisch,Wiegel} comparable in simplicity to
Eq. (\ref{eq:Spitzer}).

To our surprise, we found that this fairly old area lacks both
unity and clarity.  The studies of winding angle distribution
\cite{Yor,Privman,PitmanYor1,RudnickHu1,RudnickHu2,Duplantier_winding,PitmanYor2,Belisle,Comtet}
contain no hint on the lessons of this exactly solvable model to
polymer physics.  Drossel and Kardar \cite{Drossel} as well as
Samokhin \cite{Samokhin1,Samokhin2} brought the subject to a new
level of complexity, they examined winding angle distribution for
the random walks in a disordered medium.   Drossel and Kardar
\cite{Drossel} also provided simple derivation of the results
(\ref{eq:Spitzer},\ref{eq:NO_Spitzer}) and applied it to many
physical situations involving \emph{directed} polymers, but all
that yields little insight into the topological properties of ring
polymers.   And we are unaware of any followers of
\cite{Edwards,Prager_Frisch,Wiegel} who took advantage of the more
recent mathematical achievements
\cite{Yor,PitmanYor1,RudnickHu1,RudnickHu2,PitmanYor2,Belisle,Comtet}.
Meanwhile, an exactly solvable model in general is useful if only
it yields some insight(s).  Upon a closer look and re-examination
of the literature, we found that the model of winding can be made
to meet this criteria, but it has not been done yet. Our plan in
this paper is to re-consider the problem from a single common view
point, including both winding angle distribution and some more
physical aspects.

Our additional motivation arises from the fact that the study of
topological constraints in polymers in the years and decades after
the works \cite{Edwards,Prager_Frisch,Wiegel} had been dominated
by the phenomenological approaches based on the reptation theory
\cite{deGennesViews,DoiEdwards}.  At the same time, a breakthrough
in microscopic understanding of this subject has not been
achieved, and, therefore, the need for exactly solvable models
remains high.  Moreover, apart from networks, there is now another
large ``consumer'' for polymer topology, this is DNA physics. The
DNA double helix is frequently found in a closed loop form, it
forms knots of various kinds \cite{Cozarelli,Wang}, and there are
special enzymes spending energy to simplify the entanglements
\cite{Rybenkov}.

One of the key aspects of polymer topology is that there are two
types of questions one can ask, corresponding to \emph{annealed}
and \emph{quenched} topological disorder, respectively
\cite{Disordered_Polymers}.  The beauty of the winding model,
which so far seems to remain underappreciated, is that it allows
both types of questions:

\begin{itemize} \item  The typical annealed
topology question is that about ring closure experiment and knot
probabilities \cite{Maxim,Muthukumar,Deguchi,theorem1,theorem2}:
having a linear polymer with ``sticky'' ends, what is the
probability to obtain a certain type of a knot upon first meeting
of the two ends \cite{Cozarelli,Wang}?  A similar question for the
winding model is this: what is the probability that a random walk
on the plane links number $n$ (or winding angle $2 \pi n$) with an
obstacle?
\item
The typical quenched topology question is about, e.g., the size or
other properties of a polymer having a given fixed topology (e.g,
knot type) \cite{3_5_uzla,ExclKnot}; this is necessary, e.g., to
understand the diffusion of knotted DNA in solution or in a gel. A
s imilar question for the winding model is this: given a polymer
with fixed linking number $n$, what is the (root-mean-squared)
average distance of an arbitrary point on the trajectory from
${\cal O}$?
\end{itemize}

To conclude the introduction, we should also mention that the
shortcomings of the Edwards-Prager-Frisch model are well
understood \cite{Nechaev}.  Basically, this model assumes that
entanglements algebraically commute with each other, while the
real physical situation is non-Abelian.

This paper is organized as follows.  In section \ref{sec:Green},
we discuss the Green function formulation of the problem and
derive basic equations for all models - winding around the point,
around the disc, or inside the cavity.  In section
\ref{sec:Spitzer}, we show how to re-derive and generalize the
results (\ref{eq:Spitzer}) and (\ref{eq:NO_Spitzer}).  In section
\ref{sec:polymer}, we consider the closed loop polymer, which is
the random walk with connected ends.  In section
\ref{sec:conclusion}, we make a few final comments.


Our work is heavy on calculations, even though some of the less
important ones are relegated into Appendices.  As readers, we
don't like such heavy papers.  This is why we start from section
\ref{sec:lazy} which provides an overview of major steps and the
results for those readers not interested in details.

\section{Bird's eye view of the results: for the lazy reader who
does not want to dwell on the calculations}\label{sec:lazy}

If you, our reader, do not want to follow our calculations, this
section offers a tour of the results for you.

To begin with, section \ref{sec:Green} contains no results: it
describes the standard diffusion equation and bilinear expansion
of its Green function over the appropriate set of Bessel
functions.  Here, for a ``tourist,'' all that is necessary to know
is the notation $a^2/4$ adopted for the diffusion coefficient,
which means that the root-mean-square distance traveled by a
walker during the time $t$ is equal to $a \sqrt{t}$.

Formula (\ref{eq:Spitzer_generalized}) in section
\ref{sec:SpitzerPOINT} is our first small result, it is a very
mild generalization of the Spitzer formula (\ref{eq:Spitzer})
which takes explicit account of the distances $r$ and $r^{\prime}$
of a polymer (or random walk trajectory) ends from the origin (or
obstacle) ${\cal O}$.  Formula (\ref{eq:Spitzer_generalized})
gives the probability distribution of winding angle $\theta$ for
the random walk of length $t$ with $r$ and $r^{\prime}$ fixed.
Like the Spitzer law, it has diverging moments, such as $\langle
\theta^2 \rangle$.

The very cumbersome formula (\ref{eq:moya_veroyatnost}) gives a
similar result for the winding around a disc of a finite radius
$b$.  It generalizes formula (\ref{eq:NO_Spitzer}) by keeping
explicit track of positions $r$ and $r^{\prime}$ of both ends.
Just like (\ref{eq:NO_Spitzer}), it decays exponentially and
yields finite values for all moments, e.g., $\langle \theta^2
\rangle$. One utility of this result is the analysis of cross-over
between winding around a point with infinite $\langle \theta^2
\rangle$ and winding around a disc with finite $\langle \theta^2
\rangle$.  As we show in Section \ref{sec:Spitzer_b_to_zero}, when
the disc size $b$ goes to zero, there opens a wide range of times
$t$ (see Eq. (\ref{eq:window_of_t})) where the probability behaves
as $\sim 1 /\theta^2$ up to $\theta$ about $ \ln (a / b)$, and
only at larger $\theta$ exponential decay takes over; therefore,
when we say that $\langle \theta^2 \rangle$ diverges, this really
means $\langle \theta^2 \rangle \sim \left( \ln \left( a / b
\right) \right)^2$ at small $b$.

In Section \ref{sec:polymer} we go closer to the polymer view on
the subject. For this, we consider that the two ends of the random
walk are glued together, so that $r = r^{\prime}$ and $\theta = 2
\pi n$, where (positive or negative) integer $n$ is the linking
number, the number of turns the polymer ring makes around the
obstacle.  Figure \ref{fig:plot_W_ot_xi} depicts the statistical
weight of the polymer conformations with the linking number $n$ as
a function of $r^2/t$.  Qualitatively, this exhibits a behavior
similar to that of the knotting probability as a function of chain
length, because all cases with $n \neq 0$ (similar to non-trivial
knots) reach maximal weight at some intermediate values of $r$
and/or $t$.

Sections \ref{sec:exact_sigma} and \ref{sec:exact_r} present our
most original and most interesting findings.  In particular, we
consider the following question. Given the closed random walk
trajectory of the length $t$, we consider $\Sigma_n$ - the locus
of points around which the trajectory makes exactly $n$ turns.
Then, what is the area of $\Sigma_n$?  We denote by $\sigma_n (t)$
the average (over the random walk trajectories) of this area, then
formula (\ref{eq:inversensquare}) provides the \emph{exact} answer
to this question.  The essence is that $\sigma_n(t)$ decreases
very slowly with $n$, only as $1/n^2$.  Of course, ideologically
this is similar to the slow decay of the Spitzer distribution
(\ref{eq:Spitzer}).  Another look at the same result is to think
about a virial coefficient of a polymer ring with a long straight
bar. Their interaction is topological in nature \cite{Maxim}, and
the virial coefficient can be understood as the surface excluded
for a polymer ring by the presence of the obstacle if the ring is
not entangled with the obstacle.  This virial coefficient is the
sum of all $\sigma_n(t)$ with $n \neq 0$ , and it is
\emph{exactly} equal to $\pi t a^2 / 12$.

Note that the former view of $\sigma_n(t)$ corresponds to the
question about annealed topological disorder, as it relates
$\sigma_n(t)$ to the probability of getting the topological state
$n$.  By contrast, the latter view on the same quantities
$\sigma_n(t)$ corresponds closer to the idea of quenched
topological disorder, as it reflects on the physical property of
the polymer with given $n$.  Another such physical quantity is the
distance between the obstacle and an arbitrary point on the
polymer.  The \emph{exact} expression for the root-mean-square of
such distance is given by formula (\ref{eq:rasstoyanie}).  The
interesting aspect of this result is that this distance remains of
the order of $a \sqrt{t}$ and only quite modestly depends on $n$,
changing from approximately $0.496 a \sqrt{t}$ when $n = 1$ to
$0.408 a \sqrt{t}$ when $n \to \infty$.  The fact that this
distance decreases with growing ``topological complexity'' $n$ is
not surprising, but the fact that it changes only slightly is
interesting.  One could have thought that the polymer would
consist of $n$ roughly similar loops, leading to the typical size
of $a \sqrt{t/n}$.  Our result, therefore, suggests that even at
very large $n$ there remains one big loop, with the length of
order $t$, while all other loops are tight and small.  This is
reminiscent of knot tightening recently discussed by Kardar and
his co-workers \cite{Kardar_tightening}.

In section \ref{sec:force}, we make a brief comment on the elastic
forces developing in the polymer ring either pushed too close to
the obstacle or pulled too far away from it.

Finally, in the section \ref{sec:polymer_around_disc} we consider
polymer ring entangled with a finite size obstacle, and show that
in this case the distribution over the linking number $n$ decays
exponentially at large $n$, and the characteristic $n$ is about
$\ln ( t a^2 / b^2)$.

\section{Green function formulation}\label{sec:Green}

\subsection{Point-like obstacle}

Consider a Gaussian polymer in $2D$ or, equivalently, a random
walk in $2D$.  Suppose first that the obstacle is point-like,
positioned at ${\cal O}$, the coordinate center.  The statistics
of trajectories is fully described by the Green function, $G
\left(
\begin{array}{c} {\vec r}^{\prime} \\ 0 \end{array} \right|\left.
\begin{array}{c} {\vec r} \\ t \end{array} \right)$, which is the partition
function (or statistical weight) of the chain having the monomer
$0$ at ${\vec r}^{\prime}$ and monomer $t$ at ${\vec r}$. The
Green function satisfies the diffusion equation
\be \partial_t G \left(
\begin{array}{c} {\vec r}^{\prime} \\ 0 \end{array} \right|\left.
\begin{array}{c} {\vec r} \\ t \end{array} \right) = \frac{a^2}{4} \Delta
G \left(
\begin{array}{c} {\vec r}^{\prime} \\ 0 \end{array} \right|\left.
\begin{array}{c} {\vec r} \\ t \end{array} \right) + \delta (t) \delta({\vec r} -
{\vec r}^{\prime}) \ , \label{eq:diffusion} \ee
where the notations are standard:  $\Delta$ is the Laplace
operator acting on ${\vec r}$, $a$ is the monomer size, $t$ is
polymer length (``time'').  The notation $a^2/4$ adopted here for
the diffusion coefficient, which is in fact $a^2/2d$, $d$ being
space dimension, is convenient because root-mean-squared
end-to-end distance of the trajectory with no obstacles equals
exactly $a \sqrt{t}$.  The Green function can be written in terms
of the bi-linear expansion over the corresponding eigenfunctions.
Because our goal is to address the obstacle at ${\cal O}$, we
choose eigenfunctions with cylindrical symmetry. The ones with no
singularity at ${\cal O}$ read $J_\mu (\kappa r) e^{\pm \imath \mu
\theta}$, where $J_{\mu}(x)$ is Bessel function of the first kind,
$r$ and $\theta$ are polar coordinates corresponding to ${\vec
r}$, and $-\kappa^2$ is the corresponding eigenvalue. Accordingly,
we write
\bea G \left(
\begin{array}{c} r^{\prime},0 \\ 0 \end{array} \right|\left.
\begin{array}{c} r, \theta \\ t \end{array} \right) & = & \frac{1}{2 \pi}
\int_0^{\infty} \! \! \int_0^{\infty} e^{- a^2 \kappa^2 t / 4}
\cos ( \mu \theta ) \times \nonumber \\ & \times & J_{\mu} (
\kappa r ) J_{\mu} (\kappa r^{\prime}) \kappa d \kappa d \mu \
.\label{eq:bilinear} \eea
It is worth noting explicitly that only positive $\mu
>0$ contribute to this expansion, because $J_{\mu}(x)$ with negative index $\mu$ is singular at small $x$.

In most cases in mathematical physics, the angular
$\theta$-dependence is $2 \pi$-periodic, meaning that $\theta$ and
$\theta \pm 2 \pi n$ label one and the same place on the plane.
This is not the case for the problem at hand.  Indeed, $G_t(r,0 |
r^{\prime}, \theta)$ is the statistical weight of trajectories
(polymer conformations) that start at a point some distance $r$
away from the origin ${\cal O}$ and arrive after ``time'' $t$ at
another point some $r^{\prime}$ from ${\cal O}$, where it is
assumed that by the time $t$ the trajectory has accumulated
winding angle $\theta$ around ${\cal O}$.  Accordingly, for
instance, $\theta = 0$ means no turns around ${\cal O}$, while
$\theta = 2 \pi$ means one turn counterclockwise, $\theta = - 2
\pi$ is one turn clockwise, etc. In other words, we should treat
our plane as a Riemann surface, in which case $\theta$ and $\theta
\pm 2 \pi n$ correspond to different layers.

Most immediately, this means that not only integer, but all
positive values of $\mu$ must be included in the bilinear
expansion (\ref{eq:bilinear}).

It turns out that integration over $\kappa$ can be explicitly
performed; the derivation of the relevant so-called Weber integral
\cite{Bessel} is provided in the Appendix
\ref{sec_app:Weber_Integral}. The result reads:
\bea G \left(
\begin{array}{c} r^{\prime},0 \\ 0 \end{array} \right|\left.
\begin{array}{c} r, \theta \\ t \end{array} \right)  & = &  \frac{2}{ \pi a^2
t } e^{-( r^2 + \left.r^{\prime}\right.^2 ) / a^2 t}
 \times \nonumber \\ & \times
& \int_0^{\infty} \cos ( \mu \theta ) I_{\mu} \left( \frac{2 r
r^{\prime}}{a^2 t } \right) d \mu \ , \label{eq:afterWeber0} \eea
where $I_{\mu} (x)$ is the modified Bessel function.

It is instructive to re-write the latter formula by introducing
$R$ - the distance between ${\vec r}$ and ${\vec r}^{\prime}$:
${\vec R} = {\vec r} - {\vec r}^{\prime}$, or $R^2 = r^2 + \left.
r^{\prime} \right. ^2 - 2 r r^{\prime} \cos \theta$.  We can write
\be G \left(
\begin{array}{c} r^{\prime},0 \\ 0 \end{array} \right|\left.
\begin{array}{c} r, \theta \\ t \end{array} \right)   =  \left[ \frac{1}{ \pi a^2
t } e^{-R^2 / a^2 t}\right]  W \left(\theta, \frac{2 r
r^{\prime}}{a^2 t } \right) \ , \label{eq:afterWeber1} \ee
where
\bea W \left(\theta, \frac{2 r r^{\prime}}{a^2 t } \right) & = & 2
e^{- 2 r r^{\prime} \cos \theta / a^2 t } \times \nonumber \\ &
\times & \int_0^{\infty} \cos ( \mu \theta ) I_{\mu} \left(
\frac{2 r r^{\prime}}{a^2 t } \right) d \mu \ ,
\label{eq:afterWeber2} \eea
The first factor in the Eq. (\ref{eq:afterWeber1}) (in square
brackets) is simply the Green function of an unrestricted polymer,
or unrestricted random walk; in other words, it is the statistical
weight of all conformations going from ${\vec r}$ to ${\vec
r}^{\prime}$.  Therefore, $W$ measures the fraction of
trajectories with winding angle $\theta$ on the way.

Equations (\ref{eq:afterWeber0}-\ref{eq:afterWeber2}) were derived
by Edwards \cite{Edwards}.

\subsection{Finite size obstacle}

Consider now an obstacle having the shape of a disc with some
finite radius $b$. Since the trajectory cannot make infinitely
many turns around such obstacle, we expect that the probability
distribution for the number of turns should be completely
different for this case as compared to the point-like obstacle.

We use the same method as before.  Eq. (\ref{eq:diffusion}) still
applies, but as regards bi-liner expansion, Eq.
(\ref{eq:bilinear}), we have now different set of eigenfunctions -
the ones which satisfy the boundary condition of being equal to
zero at $r=b$. This boundary condition removes all trajectories
which cross the boundary, or, in other words, which enter the
$r<b$ region.  The eigenfunctions, corresponding to the eigenvalue
$- \kappa^2$, can be written in the form $Z_{\mu} (\kappa r ,
\kappa b) e^{\imath \mu \theta}$, where (see Appendix
\ref{appendix:Z})
\be Z_{\mu} (\kappa r , \kappa b) =  \frac{-J_{\mu} (\kappa r)
Y_{\mu} (\kappa b) + J_{\mu} (\kappa b) Y_{\mu} (\kappa
r)}{\sqrt{J_{\mu}^2 (\kappa b) + Y_{\mu}^2 (\kappa b)}}
\label{eq:Z} \ . \ee
Here $Y_{\mu}(x)$ is Bessel function of the second kind (another
frequently used notation for $Y_{\mu}(x)$ is $N_{\mu}(x)$; we
adopt here the notation used in {\em Mathematica}
\cite{Mathematica}). A few notes about functions $Z$ are provided
in the Appendix \ref{appendix:Z}, including the proof that the
square root in the denominator makes them correctly normalized.
Using $Z$, we write the Green function as a bi-linear expansion,
like Eq. (\ref{eq:bilinear}):
\bea G \left(
\begin{array}{c} r^{\prime},0 \\ 0 \end{array} \right|\left.
\begin{array}{c} r, \theta \\ t \end{array} \right) & = & \frac{1}{2 \pi}
\int_0^{\infty} \! \! \int_0^{\infty} e^{- a^2 \kappa^2 t / 4}
\cos ( \mu \theta ) \times \nonumber \\ & \times & Z_{\mu} (
\kappa r, \kappa b ) Z_{\mu} (\kappa r^{\prime}, \kappa b) \
\kappa d \kappa d \mu \ . \label{eq:bilinearZ} \eea
Unfortunately, no known analog exists of the Weber integral for
the $Z$-functions, and so, unlike the $b=0$ case above, we were
unable to find any way to simplify this by performing either of
the two integrations.

Addressing the same problem of winding around the disk, Rudnick
and Hu \cite{RudnickHu1} have already found the expression for the
Green function.  Formula (\ref{eq:bilinearZ}) looks surprisingly
different from the known result \cite{RudnickHu1}.  In the
Appendix \ref{appendix:LaplaceFourier} we show explicitly that
these two results are equivalent.

%
%

\subsection{Winding inside the cavity}

Yet another interesting model is shown in Fig.
\ref{fig:winding}(c).  It is a random walk or linear polymer
confined in a restricted volume, say, inside the disc of some
radius $B$. Then, absorbing boundary conditions should be imposed
on this boundary. Assuming for simplicity that the obstacle is
located in the center of the confinement disc, we obtain that Eq.
(\ref{eq:bilinear}) holds, except integration over $\kappa$ at
every $\mu$ must be replaced by the sum over the discrete spectrum
of $\kappa_n(\mu)$ such that $J_{\mu} (\kappa_n(\mu) B) = 0$. As
usually, as $t \to \infty$ we can resort to the ground state
dominance principle, which means we can truncate the summation to
one leading term:
\bea G \left(
\begin{array}{c} r^{\prime},0 \\ 0 \end{array} \right|\left.
\begin{array}{c} r, \theta \\ t \end{array} \right) & \simeq & \frac{1}{2 \pi}
 \int_0^{\infty} e^{- a^2 \xi_{\mu}^2 t / 4 B^2} \cos ( \mu \theta )
\times \nonumber \\ & \times & J_{\mu} ( \xi_{\mu} r /B ) J_{\mu}
(\xi_{\mu}  r^{\prime} / B) d \mu \ ,\label{eq:bilinear_cavity}
\eea
where $\xi_{\mu}$ is the smallest root of the Bessel function
$J_{\mu}(\xi)$.

\section{Winding angle distribution: Spitzer law and related
results}\label{sec:Spitzer}

\subsection{Winding around a point ($b = 0$)}\label{sec:SpitzerPOINT}

The authors of the works
\cite{Spitzer,Yor,PitmanYor1,RudnickHu1,RudnickHu2,Duplantier_winding,PitmanYor2,Belisle,Comtet},
examined the problem of winding angle distribution in the
following formulation.  Suppose the walker starts some distance
$r$ from the origin, and suppose we are interested in the winding
angle distribution irrespective of $r^{\prime}$, the distance from
the origin to the trajectory end.  Formally, such probability
distribution is obtained via suitable integration of the Green
function over $r^{\prime}$:
\be W( \theta) \propto \int_0^{\infty} G \left(
\begin{array}{c} r,0 \\ 0 \end{array} \right|\left.
\begin{array}{c} r^{\prime}, \theta \\ t \end{array} \right)  r^{\prime} d r^{\prime} \
. \label{eq:integrate_over_end} \ee
In the Appendix \ref{sec_app:integrate_over_end}, we show how to
use the Weber integral to follow this path.

Unfortunately, in some other cases considered below, such as
winding around a non-zero size disc ($b \neq 0$), we don't have
the advantage of the Weber integral simplification from Eq.
(\ref{eq:bilinear}) to Eq. (\ref{eq:afterWeber0}), which makes the
explicit integration of the Green function over $r^{\prime}$
difficult. Besides, for polymer applications it is natural to keep
track of the end position as long as possible.  This is why it is
useful to see how we can re-derive the Spitzer law
(\ref{eq:Spitzer}) directly from Eq. (\ref{eq:bilinear}), not
resorting to Eq. (\ref{eq:afterWeber0}).  This is what we shall do
now.

We note that integration over $\kappa$ in Eq. (\ref{eq:bilinear})
is effectively truncated at $\kappa^2 \leq 4/ta^2$.  When $t$ is
large enough, this leads to both $\kappa r$ and $\kappa
r^{\prime}$ being small. Then the Bessel function can be replaced
by the first term of its expansion, $J_{\mu} (\xi) \simeq
\frac{1}{\Gamma (1 + \mu ) } \left( \frac{\xi}{2} \right)^{\mu}$.
After that, the integration over $\kappa$ is easily performed,
yielding
\be G \left(
\begin{array}{c} r^{\prime},0 \\ 0 \end{array} \right|\left.
\begin{array}{c} r, \theta \\ t \end{array} \right) \simeq
\frac{1}{\pi t a^2} \int_{0}^{\infty} \left(\frac{r r^{\prime} }{t
a^2} \right)^{\mu} \frac{\cos \left( \mu \theta \right) }{\Gamma
\left( 1 + \mu \right)} d \mu \ . \label{eq:mu_malo} \ee
Assuming $r r^{\prime} / t a^2 \ll 1$ (see the discussion a few
lines below), we see that the integral over $\mu$ is dominated by
small $\mu$ in which area we can set $\Gamma \left( 1 + \mu
\right) \simeq 1$.  In this approximation, the integration over
$\mu$ is elementary, and results in
\be G \left(
\begin{array}{c} r^{\prime},0 \\ 0 \end{array} \right|\left.
\begin{array}{c} r, \theta \\ t \end{array} \right) \simeq
\frac{1}{ \pi t a^2} \frac{\ln \left( t a^2 / r r^{\prime} \right)
}{\left( \ln \left( t a^2 / r r^{\prime}\right) \right)^2 +
\theta^2} \ . \ee
%
This is the Cauchy distribution for the winding angle
\be W( \theta) = \frac{1}{\pi} \frac{1}{1+ x^2} \ ; \ x =  \frac
{\theta}{\ln \left( t a^2 / r r^{\prime} \right)}  \ .
\label{eq:Spitzer_generalized}  \ee
This result is similar, but not identical to the Spitzer formula
(\ref{eq:Spitzer}).  The difference is in the definition of the
scaling variable $x$: formula (\ref{eq:Spitzer_generalized}),
unlike (\ref{eq:Spitzer}), keeps track of the coordinates $r$ and
$r^{\prime}$ of the trajectory ends.  As we have already
mentioned, this will be useful for polymer applications.  Note,
however, that we \emph{cannot} integrate over $r^{\prime}$ as in
Eq. (\ref{eq:integrate_over_end}), because formula
(\ref{eq:Spitzer_generalized}) was derived under the assumption
that $r^{\prime}$ is not too large.

How then can we recover the Spitzer law (\ref{eq:Spitzer}) from
Eq. (\ref{eq:Spitzer_generalized})?  What we should do is to note
that one trajectory end is fixed at the distance independent of
$t$, while the other is free, meaning that $r \sim a$ and
$r^{\prime} \sim a \sqrt{t}$.  Then, we have for the scaling
quantity $x$ in formula (\ref{eq:Spitzer_generalized}) $x = \theta
\left/ \ln \left( t a^2 / r r^{\prime} \right) \right. \simeq 2
\theta / \ln t $, which is indeed exactly the same as in Eq
(\ref{eq:Spitzer}).

Other interesting extremes are as follows:  \begin{itemize}
\item If $r \sim a$ and $r \sim a t$, then the ``width'' of the
distribution gets very small. This is the closest approximation
Gaussian model can provide for the idea that fully stretched
polymer does not have any freedom to wind around the obstacle. Of
course, a Gaussian polymer cannot be fully stretched, this is why,
say, $\langle \theta^2 \rangle$, remains divergent even when the
``width'' goes to zero.  \item A similar situation is realized
when $r \sim r^{\prime} \sim a \sqrt{t}$: winding is suppressed
when the obstacle is removed to the periphery of the coil.  Note
that Eq. (\ref{eq:Spitzer_generalized}) should not be used at
larger $r$, when $\kappa r$ is not small and the Bessel function
cannot be expanded.
\item If both $r \sim a$ and $r^{\prime} \sim a$, then the result
is only different from Eq. (\ref{eq:Spitzer}) by a factor of $2$
in the definition of $x$; in this case, $x = \theta / \ln t$. That
means, fixing both ends and not allowing them to wander freely
reduces the ``width'' by half.  \end{itemize}

\subsection{Winding around a disc ($b > 0$)}\label{sec:SpitzerDISC}

For winding around a disc of finite radius $b$, we can use the
same method.  When $t$ is large enough, integration over $\kappa$
in Eq. (\ref{eq:bilinearZ}) is dominated by small $\kappa$.
Accordingly, we can resort to the small $\kappa$ expansion of
$Z_{\mu} (\kappa r, \kappa b)$ (see Eq. (\ref{eq:small_xi}) and
the discussion in the Appendix \ref{appendix:Z}):
\begin{widetext}
\be Z_{\mu} (\kappa r, \kappa b) \simeq \frac{(r/b)^{\mu} -
(r/b)^{-\mu}}{\sqrt{\left[ \left( \frac{\kappa b}{2} \right)^{\mu}
\Gamma ( 1 - \mu) - \left( \frac{\kappa b}{2} \right)^{-\mu}
\Gamma ( 1 + \mu) \right]^2 + 2 \pi \mu \tan \frac{\pi \mu }{2}}}
\ . \label{eq:expandZ} \ee
Accordingly, the $\kappa$-dependent factor in the Green function
(\ref{eq:bilinearZ}) can be presented in the form
$e^{-g(\kappa)}$, where
\be g(\kappa) = \frac{\kappa^2 t a^2}{4} + \ln \left(\left[ \left(
\frac{\kappa b}{2} \right)^{\mu} \Gamma ( 1 - \mu) - \left(
\frac{\kappa b}{2} \right)^{-\mu} \Gamma ( 1 + \mu) \right]^2 + 2
\pi \mu \tan \frac{\pi \mu }{2} \right) \ . \ee
Provided that $\mu < 1$ (which is justified a few lines below), it
is not difficult to establish that $g(\kappa)$ has a minimum,
which dominates integration over $\kappa$ at large $t$.
Straightforward differentiation yields for the the corresponding
$\kappa$ the condition
\be \kappa^2 t a^2 = 4 \mu \frac{\left( \frac{\kappa b}{2}
\right)^{-2 \mu} \Gamma^2(1 + \mu) - \left( \frac{\kappa b}{2}
\right)^{2 \mu} \Gamma^2(1 - \mu) }{\left( \frac{\kappa b}{2}
\right)^{-2 \mu} \Gamma^2(1 + \mu) + \left( \frac{\kappa b}{2}
\right)^{2 \mu} \Gamma^2(1 - \mu) - 2 \pi \mu \cot \pi \mu } \ .
\ee
\end{widetext}
This equation has just one solution which at large $t$ corresponds
to small $\kappa$.  More accurately, the solution reads
\be \frac{\kappa a}{2} \simeq \left\{ \begin{array}{lcr}
\sqrt{\frac{\mu }{ t }} & {\rm when} & \mu   \ln ( t a^2 / b^2) \gg 1   \\ \\
\sqrt{\frac{1}{t \ln (t a^2 / b^2)}} & {\rm when} & \mu \ln ( t
a^2 / b^2) \ll 1
\end{array} \right. \ . \label{eq:saddle} \ee
As it turns out, the integral over $\mu$ is dominated by $\mu \ll
1 \left/ \ln \left(  t a^2 /  b^2 \right) \right.$, so only the
lower line of the Eq. (\ref{eq:saddle}) is relevant.  For small
$\mu$, the expression for $Z_{\mu}$ Eq. (\ref{eq:expandZ}) can be
further simplified:
\be Z_{\mu} (\kappa r, \kappa b) \simeq \frac{\sinh \left( \mu \ln
( r/b) \right)}{\sinh \left( \mu \ln ( 2/\kappa b) \right)} \ .
\label{eq:expandZ1} \ee
Then, replacing $e^{-g(\kappa)}$ with its maximal value, we arrive
at the following expression for the Green function:
%
%
%
%
%
\bea G \left(
\begin{array}{c} r^{\prime},0 \\ 0 \end{array}  \right| \left.
\begin{array}{c} r, \theta \\ t \end{array} \right) & = & 
A \int_0^{\infty} d \mu \ \cos ( \mu \theta ) \times \ \ \ \ \ \ \
\ \ \ \ \ \ \ \ \ \ \ \nonumber \\ & \times & \frac{ \sinh \left(
\mu \ln \frac{r}{b} \right) \sinh \left( \mu \ln
\frac{r^{\prime}}{b} \right) }{\sinh^2 \left( \frac{\mu}{2} \ln
\frac{t a^2}{b^2} \right)} \label{eq:bilinearZ2} \eea
plus some logarithmic corrections.  In $A$ we accumulated all the
uninteresting constant prefactors, which do not depend on
$\theta$, $r$, $r^{\prime}$, and $b$.

Now, considering this integral over $\mu$, we have to justify all
the assumptions and approximations which we made on the way. First
and foremost, the assumption that $\mu$ is small is justified by
the rapid convergence of the integral (\ref{eq:bilinearZ2}).
Indeed, at large $\mu$ all three $\sinh$'s can be replaced by
positive exponentials, leaving us with $\exp\left[- \mu \left( \ln
\frac{t a^2}{b^2} - \ln \frac{r}{b} - \ln \frac{r^{\prime}}{b}
\right) \right]$. Since $rr^{\prime} \ll t a^2$, the latter two
logarithms in the round brackets should be neglected.  That means,
the convergence of the integral (\ref{eq:bilinearZ2}) is
controlled by the $\sinh$ in the denominator, which effectively
truncates integration at $\mu$ smaller than $1/ \ln \left( t a^2 /
 b^2 \right)$.  This is very good news.  First of all, since $1/
\ln \left( t a^2 /  b^2 \right) \ll 1$, this justifies the small
$\mu$ simplification performed in formula (\ref{eq:expandZ1}).
Second of all, this also justifies the use of the lower line in
the expression (\ref{eq:saddle}) for the saddle point. Third,
since only small $\mu$ contribute, the validity condition for the
expansion of Bessel functions in the first step of Eq.
(\ref{eq:expandZ}), which generally reads $\left(\kappa r
\right)^2 \ll 1 + \mu$, can be simplified to $r^2 \ll a^2 t$ (and
similarly $\left. r^{\prime}\right.^2 \ll a^2 t$).

Thus, all approximations leading to the expression
(\ref{eq:bilinearZ2}) are self-consistent.  The only task left is
to evaluate the integral (\ref{eq:bilinearZ2}). This task gets
easier if we use the notations
\be \alpha = \frac{2 \pi \ln \frac{ r }{ b } }{\ln \frac{t a^2
}{b^2 }} \ , \ \alpha^{\prime} =  \frac{2 \pi \ln \frac{
r^{\prime} }{ b } }{\ln \frac{t a^2 }{b^2 }} \ , \ x = \frac{2
\theta }{\ln \frac{t a^2 }{b^2 }} \ . \label{eq:scaling_variables}
\ee
Then, formula (\ref{eq:bilinearZ2}) is transformed into the
following expression for the probability distribution of the
winding angle $\theta$, or, better, of the scaling variable $x$,
at fixed $r$ and $r^{\prime}$:
\be W(\theta)  =  \frac{\pi}{2 \alpha \alpha^{\prime}}
\int_{-\infty}^{+\infty} \frac{\sinh \left (\xi \alpha /\pi
\right) \sinh \left (\xi \alpha^{\prime} /\pi \right)}{\sinh^2
\xi} e^{\imath x \xi} d \xi \ , \ee
where we have re-introduced the normalization factor, such that
$\int_{-\infty}^{+\infty} W(\theta) d x = 1$.  This integral
can be reduced to the infinite sum of residues corresponding to
the poles along the imaginary axes on the complex $\xi$-plane.  In
turn, the resulting sum (which is the combination of several
geometric series) is easy for \emph{Mathematica}
\cite{Mathematica}, but can be also computed by hand.  One way or
the other, here is the result:
\begin{widetext}
%
%
\be  W(\theta) =  \frac{\pi}{2 \alpha \alpha^{\prime}
} \frac{\pi x \sinh \pi x \ \sin \alpha \ \sin \alpha^{\prime} +
\alpha \sin \alpha^{\prime} \ \left[\cos \alpha^{\prime} - \cos
\alpha \ \cosh \pi x \right]+ \alpha^{\prime} \sin \alpha \
\left[\cos \alpha - \cos \alpha^{\prime} \ \cosh \pi x \right]
}{\left[ \cosh \pi x - \cos \left( \alpha - \alpha^{\prime}
\right) \right]\times \left[ \cosh \pi x - \cos \left( \alpha +
\alpha^{\prime} \right) \right] } \ . \label{eq:moya_veroyatnost}
 \ee
\end{widetext}
The result (\ref{eq:moya_veroyatnost}) is unfortunately quite
cumbersome, although it is symmetric and in some ways quite nice.
Its beauty is revealed by consideration of various limits.  As we
learned in the case of point-like obstacle, the most interesting
limit is when chain end is free, meaning that $r^{\prime} \sim a
\sqrt{t}$.  Then, provided only that $t a^2 \gg b^2$ - which is
necessary, as the walker must have traveled much farther than the
obstacle size $b$, we get $\alpha^{\prime} \simeq \pi$.  In this
case, we get
\be W (\theta) = \frac{\pi}{2} \frac{\sin \alpha / \alpha}{\cosh
\pi x + \cos \alpha } \stackrel{\alpha \to 0}{\longrightarrow}
\frac{\pi / 4}{\cosh ^2 \left( \pi x / 2 \right)} \ , \ee
where in the latter transformation we also noted that as the
trajectory starting point is fixed, $r$ is independent of $t$, or
$\alpha \to 0$ at large $t$.  Thus, we recover formula
(\ref{eq:NO_Spitzer}). Importantly, the definition of scaling
variable $x$ (\ref{eq:scaling_variables}) becomes identical to
that in (\ref{eq:NO_Spitzer}), again under the same condition $t
a^2 \gg b^2$.

As in the case of point-like obstacle, other interesting extremes
are as follows: \begin{itemize} \item Both $r \sim a \sqrt{t}$ and
$r^{\prime} \sim a \sqrt{t}$.   In fact, this case is on the
border of applicability of our approximations, but qualitatively
the result holds. Indeed, $W(\theta)$ becomes very narrow, and
approaches $\delta (x)$. This means, no turns are possible around
the obstacle which is away from the random walk trajectory.
\item  Another case, and also a border case in terms of applicability
of our approximations, is $r \sim a t$ or $r^{\prime} \sim a t$,
implying an exponentially improbable straight trajectory.  The
distribution is again sharply localized at small $x$.  \item  Both
$r$ and $r^{\prime}$ are independent of $t$, meaning that both
$\alpha \simeq 0$ and $\alpha^{\prime} \simeq 0$.  This case is
safely within the limits of applicability.  Then,
\be W(\theta) = \frac{\pi}{2}\frac{\pi x \sinh \pi x + 2 \left( 1
- \cosh \pi x \right) }{\left( 1 - \cosh \pi x \right)^2} \ .
\label{eq:dva_konca_fixed} \ee
As in the $b=0$ case, this distribution, as one can easily check,
is exactly two times more narrow than that of
(\ref{eq:NO_Spitzer}).  In this case, unlike $b=0$, this statement
can be formalized by looking at the second moment of the
distributions (\ref{eq:NO_Spitzer}) and
(\ref{eq:dva_konca_fixed}), which (in terms of $x$) turns out
equal $1/3$ and $2/3$, respectively.
\end{itemize}

\subsection{$b \to 0$ limit: applicability conditions of the Spitzer formula}\label{sec:Spitzer_b_to_zero}

According to the Eqs. (\ref{eq:Spitzer}) and
(\ref{eq:NO_Spitzer}), winding angle distribution has finite
variance at $b \neq 0$ and diverging infinite variance at $b=0$.
These equations leave it unclear what happens when the obstacle
gets smaller and smaller, or when $b$ decreases and approaches
$0$.  It is instructive and interesting to use formula
(\ref{eq:moya_veroyatnost}) to see what really happens when $b \to
0$.

The important part of our analysis here is to realize that so far
we have been using several differently defined scaling variables
$x$: see Eqs. (\ref{eq:Spitzer}), (\ref{eq:NO_Spitzer}),
(\ref{eq:Spitzer_generalized}), (\ref{eq:scaling_variables}).  So
far, it was (hopefully) clear from the context in every place
which $x$ we have in mind.  Now, when we examine the $b \to 0$
limit, we shall face several of these different $x$
simultaneously, so we must be certain as to which $x$ is which.
For the rest of this section, we adopt the notation in which each
$x$ is labeled with the number of the defining equation:
$x_{\ref{eq:Spitzer}}$, $x_{\ref{eq:Spitzer_generalized}}$,
$x_{\ref{eq:scaling_variables}}$ (note, that
$x_{\ref{eq:NO_Spitzer}}$ is exactly the same as
$x_{\ref{eq:Spitzer}}$: $x_{\ref{eq:Spitzer}} \equiv
x_{\ref{eq:NO_Spitzer}}$).  In particular, $x$ in formula
(\ref{eq:moya_veroyatnost}) is, of course,
$x_{\ref{eq:scaling_variables}}$.

Speaking of different definitions of $x$, we should realize that
so far we have been presenting probability distributions
$W(\theta)$ normalized with respect to integration over the
corresponding $x$.  For our purposes now, it is more convenient to
use the normalization condition with respect to angle $\theta$:
$\int_{-\infty}^{\infty} W(\theta) d \theta =1$.  For the formula
(\ref{eq:moya_veroyatnost}), this means the factor $2 / \ln (t a^2
/ b^2)$ should be incorporated; we do not re-write the formula for
brevity.

The main reason why the difference between various $x$ was
unimportant so far is that at $t \to \infty$ all definitions
converge to the same: $x_{\ref{eq:Spitzer_generalized}} \to
x_{\ref{eq:scaling_variables}} \to x_{\ref{eq:Spitzer}} = 2 \theta
/ \ln t$.  However, when $b \to 0$, there appears a very broad
intermediate range of times $t$ such that although $t a^2 \gg b^2
$ (the trajectory is long enough to wind around the obstacle), but
$ t b^2 \ll a^2$:
\be a^2 / b^2 \gg t \gg b^2/a^2 \ . \label{eq:window_of_t} \ee
This is the range which we must examine.  In this range, to the
leading approximation, $x_{\ref{eq:scaling_variables}}$ does not
depend on time:
\be x_{\ref{eq:scaling_variables}} = \frac{2 \theta }{\ln \frac{t
a^2 }{ b^2 }} \simeq \frac{\theta}{\ln (a/b)} \ .
\label{eq:x_is_small} \ee
Furthermore, there is a broad range of winding angle $\theta$ in
which $x_{\ref{eq:scaling_variables}}$ is small.

Now, we should look at the quantities $\alpha$ and
$\alpha^{\prime}$.  When $b \to 0$, both of them turn out to be
slightly below $\pi$.  For instance,
\be \alpha = \frac{2 \pi \ln \frac{ r }{ b } }{\ln \frac{t a^2
}{b^2 }}  =  \pi \frac{\ln  \frac{a}{b} + \ln \frac{r}{a} } {\ln
\frac{a}{b} + \frac{1}{2} \ln t} \equiv \pi - \delta \ ,
\label{eq:delta_is_small} \ee
where
\be \delta  =  \frac{\ln \frac{t a^2}{r^2} }{2 \ln \frac{a}{b}}
\ll 1 \ .  \ee
Similarly, $\alpha^{\prime} = \pi - \delta^{\prime}$, with
similarly defined $\delta^{\prime} \ll 1$.

Thus, we can simplify formula (\ref{eq:moya_veroyatnost})
resorting to expansion of both numerator and denominator over the
powers of $x_{\ref{eq:scaling_variables}}$, $\delta$, and
$\delta^{\prime}$.  In fact, as we see from Eqs.
(\ref{eq:x_is_small},\ref{eq:delta_is_small}) all these expansions
are ones over the inverse powers of $\ln (a/b)$, and we keep the
leading terms only.  Incorporating, as explained above, the factor
$2 / \ln (t a^2 / b^2) \simeq 1/ \ln (a /b)$ to establish the
normalization $\int_{-\infty}^{\infty} W(\theta) d \theta =1$, we
finally get
\begin{widetext}
\bea W(\theta) & \simeq & \frac{1}{\ln \frac{a}{b}} \times
\frac{\frac{\pi^2}{2} \left( \delta - \delta^{\prime} \right)^2
\left( \delta + \delta^{\prime} \right) + \frac{\pi^4}{2} \left(
\delta + \delta^{\prime} \right) x_{\ref{eq:scaling_variables}}^2
}{\frac{\pi^2}{2} \left( \delta - \delta^{\prime} \right)^2 \left(
\delta + \delta^{\prime} \right)^2 + \frac{\pi^4}{2} \left[ \left(
\delta + \delta^{\prime} \right)^2 + \left( \delta -
\delta^{\prime} \right)^2 \right] x_{\ref{eq:scaling_variables}}^2
+ \frac{\pi^6}{2} x_{\ref{eq:scaling_variables}}^4 } = \nonumber
\\ & = &
\frac{1}{\pi} \frac{\underbrace{\left( \ln \frac{r^{\prime}}{r}
\right)^2 \ln \frac{t a^2}{r r^{\prime}}} + \ln \frac{t a^2}{r
r^{\prime}} \theta^2}{\underbrace{\left( \ln \frac{r^{\prime}}{r}
\right)^2 \left( \ln \frac{t a^2}{r r^{\prime}} \right)^2} +
\left[ \underbrace{\left( \ln \frac{r^{\prime}}{r} \right)^2} +
\left( \ln \frac{t a^2}{r r^{\prime}} \right)^2 \right] \theta^2 +
\theta^4 } \simeq \frac{1}{\pi} \frac{ \ln \frac{t a^2}{r
r^{\prime}} }{
 \left( \ln \frac{t a^2}{r r^{\prime}} \right)^2  + \theta^2 } \ .
\eea
\end{widetext}
For the two latter steps, we have plugged in the explicit
expressions for $\delta$, $\delta^{\prime}$
(\ref{eq:delta_is_small}), and $x_{\ref{eq:scaling_variables}}$
(\ref{eq:x_is_small}), and then neglected the $\delta -
\delta^{\prime} \sim \ln (r^{\prime} /r)$ terms (which we have
underbraced in the intermediate formula).  The result is exactly
the same as formula (\ref{eq:Spitzer_generalized}) (except it is
normalized with respect to integration over $\theta$).

From our analysis, we can now understand the cross-over between
Eqs. (\ref{eq:NO_Spitzer}) and (\ref{eq:Spitzer}). Specifically,
the Spitzer formula (\ref{eq:Spitzer}) and its generalization
(\ref{eq:Spitzer_generalized}) apply as long as two conditions are
met:  $t \ll a^2 / b^2$ and $\theta \ll \ln (a/b)$. At longer
times and/or larger angles, the exponential tail of the
distribution takes over.  For instance, when we say the $\langle
\theta^2 \rangle$ diverges for winding around a very small
obstacle, this really means $\langle \theta^2 \rangle \sim \left(
\ln ( a / b) \right)^2$.

\subsection{Winding inside a cavity}

We start from Eq. (\ref{eq:bilinear_cavity}).  It is not difficult
to realize that $\xi_\mu$ (the smallest zero of $J_{\mu} (\xi)$)
increases with $\mu$.  Therefore, when $t$ gets large, the
integration over $\mu$ is dominated by small $\mu$, as in all
previous cases.  At small $\mu$, $\xi_{\mu}$ is a smooth
non-singular function, we can linearize it: $\xi_{\mu} \simeq
\xi_{0} + \mu \xi_0^{\prime}$.  Numerically, $\xi_0 \approx 2.405$
and $\xi_0^{\prime} \approx 1.543$.  To the same approximation,
$J_{\mu}(\xi_{\mu} r / B) \simeq J_0 (\xi_0 r / B)$.  Therefore,
evaluation of the integral in Eq. (\ref{eq:bilinear_cavity})
becomes trivial, and the result reads
\be G \left(
\begin{array}{c} r^{\prime},0 \\ 0 \end{array} \right|\left.
\begin{array}{c} r, \theta \\ t \end{array} \right)  \simeq  \frac{1}{2}
J_{0} \left(\frac{\xi_{0}  r }{B} \right) J_{0}
\left(\frac{\xi_{0}  r^{\prime} }{B} \right) W (\theta) \ ,
\label{eq:otvet_cavity} \ee
where the probability distribution of the winding angle is given
by
\be W(\theta) = \frac{1}{\pi} \frac{ta^2 \xi_0 \xi_0^{\prime} / 2
B^2 }{\left( ta^2 \xi_0 \xi_0^{\prime} / 2 B^2 \right)^2 +
\theta^2 } \ . \label{eq:otvet_cavity2} \ee

The decoupling of the ends $r$ and $r^{\prime}$ in formula
(\ref{eq:otvet_cavity}) is not surprising, this is the property of
random walk locked in a restricted volume, and it is due to the
fact that correlations are broken every time that the trajectory
is reflected from the cavity border.  As regards probability
distribution of winding angle, it is once again the Cauchy
distribution, however, the scaling variable involves $\theta /
\sqrt{t}$ instead of $\theta / \ln t$ for the random walk in an
unrestricted space.  This is also because correlations are broken
every time that the trajectory hits the border.  One can say that
pieces of random walk with length about $(B/a)^2$ act
independently of each other.

This gives rise to the following simple scaling argument providing
an insight into the result (\ref{eq:otvet_cavity}).  The winding
angle distribution for every ``blob'' of the length $\sim (B/a)^2$
is given by the Spitzer formula (\ref{eq:Spitzer}), with the
replacement $t \to (B/a)^2$.  Now, we have $t / (B/a)^2$ of such
blobs.  Since blobs are independent, the probability distribution
of the sum of all winding angles of all blobs is given as a
convolution.  In other words, Fourier transform of the Spitzer
distribution for one blob, which is $e^{- \left| \mu \right| \ln
(B/a)^2}$, must be taken to the power $t / (B/a)^2$.  Apart from
logarithmic corrections, this returns the result
(\ref{eq:otvet_cavity}).

Thus, the reason  why $\langle \theta^2 \rangle$ diverges for the
polymer inside the cavity is because every blob can make many
turns around the point-like obstacle on a small scale, before ever
hitting the border of the cavity.

\section{Ring polymer: Edwards-Prager-Frisch model}\label{sec:polymer}

We now want to make one step closer to the attempt of gaining
insight into the properties of closed ring polymers.  One way in
this direction would be to say that a ring polymer is the random
walk trajectory whose end points happen to coincide, namely $r =
r^{\prime}$ and $\theta = 2 \pi n$, where integer $n$ (positive,
negative, or zero) is the linking number (number of turns).  Our
results
(\ref{eq:Spitzer_generalized},\ref{eq:moya_veroyatnost},\ref{eq:dva_konca_fixed},\ref{eq:otvet_cavity2})
are suitable for this, and we shall do it.  It turns out also
useful, however, to derive some additional results independently.
In particular, some of the results below are exact (not
asymptotically exact, but just exact).

\subsection{Point-like obstacle: a ring with one monomer anchored}

Thus, we return to Eqs.
(\ref{eq:afterWeber1},\ref{eq:afterWeber2}), and use them this
time to write down the statistical weight of the ring polymer
conformations with linking number $n$ and with one monomer fixed
at the distance $r$ from ${\cal O}$:
\bea G_n (r,t) & \equiv & G \left(
\begin{array}{c} r,0 \\ 0 \end{array} \right|\left.
\begin{array}{c} r, 2 \pi n \\ t \end{array} \right)   = \frac{1}{\pi
a^2 t }   W_n ( \xi ) \ ;  \\
W_n ( \xi ) & = &  2 e^{- \xi } \int_0^{\infty} \cos ( 2 \pi n \mu
) \ I_{\mu} ( \xi ) \ d \mu \ ; \ \  \xi = \frac{2 r^2}{a^2 t } \
. \nonumber   \label{eq:Wreturn} \eea
In this formula, $G_n (r,t)$ is the statistical weight of the ring
with $n$ turns, while the prefactor $1/ \pi a^2 t$ is the
statistical weight of a ring with no topological constraints
\footnote{Using the notion of statistical weight, one has to be
conscious of the issue of distinguishability of monomers.  For
instance, if all of them are exactly identical, then extra factor
$t$ must be incorporated in the statistical weight of a ring.
Here, we assume that since one monomer is anchored, the monomers
are distinguishable by their numbering along the chain starting
from the anchored place.  It is also not difficult to establish
that difficulties do not arise later, in Section
\protect\ref{sec:exact_sigma} when we consider a ring with no
anchored monomer. We are indebted to I. Ya. Erukhimovich for this
important comment.}. Therefore, $W_n ( \xi )$ is the probability
that polymer ring fixed at one point $r$ makes $n$ turns around
the obstacle. In the Appendix \ref{sec_app:normalization_of_W} we
check explicitly that $W_n$ satisfies the normalization condition
as a probability.

Similarly to what we did before, we can address the case $t a^2
\gg r^2$, or $\xi \ll 1$.  In this case, we truncate the small
$\xi$ expansion of $I_{\mu}(\xi) \simeq \left( \xi / 2
\right)^{\mu} / \Gamma ( 1 + \mu )$, replace $\Gamma ( 1 + \mu )
\simeq 1$ (compare Eq.
(\ref{eq:mu_is_small_when_integr_over_end})), and then obtain
\be W_n (\xi ) \simeq 2 (1-\xi)  \frac{\ln (2/\xi) }{\left( \ln (
2 / \xi ) \right)^2 + 4 \pi^2 n^2 } \ , \ \ \ \xi \to 0 \ .
\label{eq:Spitzer_Discrete} \ee
Of course, this is nothing else but the ``discrete'' version of
the Spitzer distribution.  However, merely taking $\theta = 2 \pi
n$ in Eq. (\ref{eq:Spitzer_generalized}) is not enough, as the
normalization factor in (\ref{eq:Spitzer_generalized}) corresponds
to $\int_{-\infty}^{+\infty} W (\theta) d \theta = 1$, while in
Eq. (\ref{eq:Spitzer_Discrete}) it corresponds to
$\sum_{n=-\infty}^{+\infty} W_n (\xi) = 1$.

For a polymer, it makes perfect sense to examine also the opposite
extreme, $\xi \gg 1$.  The corresponding asymptotics are easy to
derive from the somewhat simplified expression for $W_n(\xi)$.

We can afford further simplification of the expression Eq.
(\ref{eq:Wreturn}) for $W_n (\xi)$ resorting to the following
integral representation of the modified Bessel function \cite{GR}:
\bea I_{\mu}( \xi) & = &  \frac{1}{2 \pi} \int_{- \pi}^{\pi}
e^{\xi \cos \beta} \cos ( \beta \mu ) d \beta - \nonumber \\ & - &
\frac{\sin ( \mu \pi )}{\pi} \int_0^{\infty} e^{- \xi \cosh u -
\mu u} du \ , \eea
which is generally valid at $\left| {\rm Arg} \xi \right| \leq \pi
/ 2$ and $\Re \mu > 0$.  Both of these conditions are met in our
case.  Substituting this into the Eq. (\ref{eq:Wreturn}), one can
easily perform the integration over $\mu$ yielding
\bea \label{eq:singleintegral} W_n ( \xi ) & = & \Delta_{n0} +
\int_0^{\infty} d u \ e^{- \xi \left( 1 + \cosh u \right)} \times
\\ & \times & \left[ \frac{2n-1}{u^2 + \pi^2 \left( 2 n -1
\right)^2} - \frac{2n + 1}{u^2 + \pi^2 \left( 2 n + 1
\right)^2}\right]  \ , \nonumber
 \eea
where $\Delta_{n0}$ is the Kronecker symbol ($1$ for $n=0$ and $0$
otherwise).  The latter result for $n=0$ is worth re-writing
separately:
\be W_0 ( \xi )  =  1 - 2 \int_0^{\infty}  \ \frac{e^{- \xi \left(
1 + {\rm \cosh} u \right)}}{u^2 + \pi^2 } \ du  \ .
\label{eq:Phi0} \ee

\begin{figure}[ht]
\centerline{\scalebox{0.6} {\includegraphics{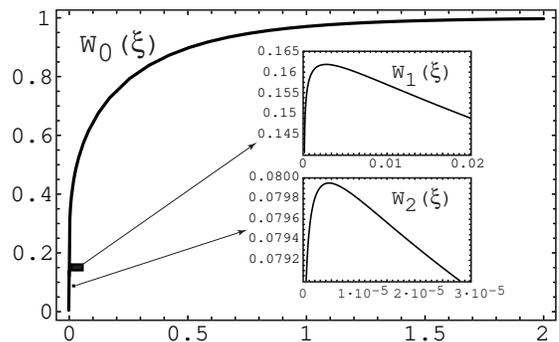}}}
\caption{ $W_n(\xi)$ is the probability to form a link of order
$n$ with the point-like obstacle provided that one of the chain
points is fixed at the distance $r$ away from the obstacle, where
$\xi = 2 r^2 / a^2 t$, $t$ being the chain length.  The plots
present the result of numerical integration based on formula
(\protect\ref{eq:singleintegral}).   The plot of $W_0(\xi)$ is
presented in the main figure.  $W_n(\xi)$ with $n >0$ would not be
seen well in this scale.  For both $W_1(\xi)$ and $W_2(\xi)$, we
show the inset, each presenting the vicinity of the maximum; the
corresponding places on the main figure are shown by tiny dark
rectangles. However small may seem every particular $W_n(\xi)$, it
should be born in mind that together they sum up to $1 -
W_0(\xi)$. }\label{fig:plot_W_ot_xi}
\end{figure}

Equations (\ref{eq:singleintegral},\ref{eq:Phi0}) are convenient
enough to address the $\xi \gg 1$ extreme.  Indeed, when $\xi$ is
large, the integral converges at small $u$, which allows us to
neglect $u$ everywhere except in the exponential factor, where we
can also truncate the $\cosh u \simeq 1 + u^2 /2$.  This yields:
\be W_n (\xi) \simeq  \cases{  \frac{\sqrt{2 / \pi^3}}{4 n^2 -1}
\frac{e^{-2 \xi}}{\sqrt{\xi}}, & $n \neq 0$ \cr 1- \sqrt{2 /
\pi^3} \frac{e^{-2 \xi}}{\sqrt{\xi}}, & $n=0$  \cr } \ \ \xi \to
\infty \label{eq:largexi} \ee
Thus, comparing (\ref{eq:Spitzer_Discrete}) and (\ref{eq:largexi})
all $W_n(\xi)$ start at $0$ at $\xi =0$ and grow very rapidly at
small $\xi$.  At $n=0$, $W_0(\xi)$ keeps increasing monotonically
with $\xi$, and $W_0(\xi)$ approaches the saturation level of $1$,
while all $W_n(\xi)$ with $n \neq 0$ decrease and rapidly die away
at large $\xi$. Obviously, each of them goes through a maximum. It
is not difficult to establish that the maximum of $W_n(\xi)$
corresponds to $\xi \sim 1/ \cosh ( \pi \sqrt{4 n^2 - 1} )$ which
at large $n$ corresponds to $\xi \sim e^{- 2 \pi n}$. This is
consistent with the fact that small $\xi$ asymptotics Eq.
(\ref{eq:Spitzer_Discrete}) is valid at $\xi \ll e^{- 2 \pi n}$.

Eq. (\ref{eq:singleintegral}) allows also straightforward
numerical integration which results in the plots shown in the Fig.
\ref{fig:plot_W_ot_xi}.

\subsection{Ring polymer entangled with a point}\label{sec:exact_sigma}

For the ring polymer, it is not very natural to consider one
monomer being fixed at $r$;  all monomers of a ring are
equivalent. Accordingly, it is natural to define the quantity
\be \sigma_n (t)  = \int_0^{\infty} \! W_n( \xi) 2 \pi r dr =
\frac{\pi a^2 t}{2} \int_0^{\infty} \! W_n(\xi) d \xi \ .
\label{eq:defsigma} \ee
What is $\sigma_n (t)$? This quantity has the units of surface
area and can be interpreted in the following way. Suppose a ring
polymer moves freely on the plane within some large area $A$ (much
larger than the polymer size, so the polymer is not restricted in
terms of its conformation). Consider one particular conformation
of our polymer and then choose a random point ${\cal O}$ within
$A$. Then $\sigma_n/A$ is the probability that polymer makes an
$n$-fold link around ${\cal O}$.  We expect physically that
$\sigma_0$ should be large, almost as large as $A$.  This is also
seen directly from the Eq. (\ref{eq:Phi0}): when integrated over
the whole area $A$, the first term (unity) yields just $A$.   This
is because when ${\cal O}$ is outside the coil, there may not be
any topological links.  If and only if the random point ${\cal O}$
is located within the polymer coil can there be any topological
link. Therefore, $\sigma_0$ should be less than $A$ by a quantity
of the order of the coil gyration radius squared, which is of the
order $t a^2$.  On the other hand, $\sigma_n$ with $n \neq 0$
should be themselves of order $t a^2$, or even smaller.

Another way of understanding of $\sigma_n(t)$ is this.  Consider
one particular conformation of a ring and consider the set of
points $\Sigma_n$ such that the polymer makes linking with $n$
turns around every point of $\Sigma_n$.  Then, $\sigma_n(t)$ is
the surface area, or the measure, associated with the set
$\Sigma_n$.

Trying to compute $\sigma_n (t)$, we can resort to either of the
expressions (\ref{eq:Wreturn}) or (\ref{eq:singleintegral}).  Let
us first explore the first possibility:
\bea \sigma_n (t)  = \pi t a ^2 \int _0^{\infty} e^{- \xi}
\int_0^{\infty} \cos ( 2 \pi n \mu ) I_\mu (\xi) d \mu d \xi \ .
\eea
Here, we face a difficulty, because the integral
\be \int_0^{\infty} e^{-\xi} I_{\mu} (\xi) d \xi
\label{eq:diverges} \ee
\emph{diverges} at large $\xi$.  What is the physical meaning of
this? Of course, this is because $\sigma_0(t)$ is close to $A$,
or, in other words, it is divergent unless we take into account
overall volume restriction.  We conclude, therefore, that the
integral (\ref{eq:diverges}) diverges for a good reason:  this is
because unlinked polymer is free to move away from the obstacle,
making $\sigma_0$ as large as (almost) $A$.

This hints on the way to circumvent the problem.  Let us assume
that the polymer is attached to the point ${\cal O}$ by a very
weak spring.  Since such polymer does not move away even when
there are no topological links, we expect that even $\sigma_0$
should remain finite, independent of $A$.  Indeed, instead of
(\ref{eq:diverges}) we have now
\be \int_0^{\infty} e^{- \alpha  \xi} I_{\mu} (\xi) d \xi =
\frac{1}{\sqrt{\alpha^2 - 1} \left( \alpha  + \sqrt{\alpha^2 -1}
\right)^{\mu}} \label{eq:converges} \ee
which \emph{converges} at any $\alpha > 1$; here $\alpha -1$ is
the effective spring constant.  Of course, we will take $\alpha
\to 1$ at the end.  Performing the remaining integration over
$\mu$, we arrive at
\be \sigma_n (t, \alpha) = \frac{\pi t a^2  \ln \left( \alpha +
\sqrt{\alpha^2 -1} \right) }{\sqrt{\alpha^2 - 1} \left[ ( 2 \pi n
)^2  + \left( \ln \left( \alpha + \sqrt{\alpha^2 -1}\right)
\right)^2 \right]} \label{eq:dopredela}\ee
As expected, the $\alpha \to 1$ limit can now be performed with no
difficulties at every $n \neq 0$, yielding finally
\be \sigma_n (t) = \frac{ t a^2}{ 4 \pi} \frac{1}{ n^2} \ , \ \ \
n \neq 0 \ . \label{eq:inversensquare} \ee
Accordingly,
\be \sigma_0 (t) = A - \sum_{n \neq 0} \sigma _n (t) = A - \frac{
\pi }{12} t a^2 \ . \label{eq:sigma0} \ee
That result exactly can be also obtained plugging Eq.
(\ref{eq:singleintegral}) into the Eq. (\ref{eq:defsigma}),
although, somewhat surprisingly, calculations are more involved
along this route.

We would like to remind to our reader once again, that $\sigma_n
(t) / A$ is the probability to have linking number $n$ for the
polymer of the length $t$.  As the probability distribution,
$\sigma_n$ has the peculiarity that all its moments obviously
diverge, even just the average linking number is infinite.  It is
not difficult to trace this back to the fact that infinitely
flexible polymer, as represented by the Brownian random walk
trajectory, can make infinitely many turns around a point-like
obstacle.  We shall address this further later.

It is worth emphasizing that the results
(\ref{eq:inversensquare},\ref{eq:sigma0}) are exact, their
validity does not require even that $t$ is large - they are exact
at any $t$.

\subsection{How far is the ring from the point-like
obstacle?}\label{sec:exact_r}

One more interesting quantity to look at is $\langle r^2 \rangle$:
the mean squared distance of one particular point on the ring to
the obstacle, ${\cal O}$.  To determine the probability
distribution for $r$, we note that $\sigma_n(t)$ plays the role of
a partition function.  The probability density for $r$ reads
\be \frac{ G \left(
\begin{array}{c} 0,r \\ 0 \end{array} \right|\left.
\begin{array}{c} 2 \pi n, r \\ t \end{array} \right)}{\int_0^{\infty} G \left(
\begin{array}{c} 0,r \\ 0 \end{array} \right|\left.
\begin{array}{c} 2 \pi n, r \\ t \end{array} \right) 2 \pi r d r} = \frac{W_n ( \xi) }{\sigma_n (t)} \ . \ee
When computing $\langle r^2 \rangle$ from this, formula
(\ref{eq:dopredela}) comes in handy, as $\langle r^2 \rangle$ is
basically the derivative of $\sigma_n (t, \alpha)$ with respect to
$\alpha$ at $\alpha = 1$:
\bea \langle r^2 \rangle & = & \frac{\pi}{\sigma_n(t)} \left(
\frac{a^2 t}{2} \right)^2 \int_0^{\infty} W_n ( \xi) \xi d \xi =
\nonumber \\ & = & - \left. \frac{ a^2 t}{2 \sigma_n (t)}
\frac{\partial \sigma_n (t, \alpha)}{\partial \alpha}
\right|_{\alpha = 1} \ . \eea
Straightforward calculation yields
\be \langle r^2 \rangle  =  \frac{a^2 t}{6} \left[ 1 + \frac{3}{2
\pi n^2} \right] \ . \label{eq:rasstoyanie} \ee
The result is interesting.  Surprisingly, it goes to a finite
constant proportional to the unperturbed coil size $ta^2$ in the
limit of very strong linking, $n \to \infty$.  This should be
understood by noting that even very large number of turns will be
produced by a short piece of a trajectory, leaving a long part, of
the order $t$, unentangled, with the size of order $a^2t$.

This is reminiscent of the recent findings by Kardar and his
co-workers \cite{Kardar_tightening} in which they claim that in
many cases real knots in three dimensional polymeric loops are
entropically dominated by conformations with the knot tightened in
a short piece of polymer, and with the rest of the polymer
fluctuating freely with no knots.

\subsection{Force}\label{sec:force}

When winding model was first introduced in the polymer physics
\cite{Edwards,Prager_Frisch,Wiegel}, it was done mostly in
connection with problems related to the rubber elasticity.
Accordingly, elastic force, or force-extension curve, was the
primary subject of interest. In case of DNA, such a curve can be
also measured using some sort of a single molecule technique
\cite{Quake}.  Although both in rubbers and in DNAs real forces
have both entropic and enthalpic contributions, in the winding
model the force has purely entropic nature and, therefore, it is
proportional to $k_B T$ in standard notations, where $k_B$ is
Boltzmann constant and $T$ is absolute temperature.  In our
notations, the force $f_n$ which should be applied to the polymer
to keep one of its links a certain distance $r$ from the obstacle
${\cal O}$ at the fixed topological invariant $n$ is given by
\be \frac{f_n}{k_B T}  = - \frac{\partial \ln W_n}{\partial r} = -
\frac{4 r}{t a^2 W_n} \frac{\partial W_n}{\partial \xi} \ . \ee
We have not found any simple closed expression for the force, in
this sense we make really no progress on this point compared to
the papers \cite{Edwards,Prager_Frisch,Wiegel}.  Nevertheless,
qualitatively, one glance at the Figure \ref{fig:plot_W_ot_xi} is
sufficient to realize that the force $f_0$ of an unentangled ring
is always positive.  This is obviously because this ring is
topologically repelling the obstacle.  On the other hand, when $n
\neq 0$, the force is positive, corresponds to repulsion, only
when $r$ (or $\xi$) is small enough.  At larger $r$ (or $\xi$),
the force flips sign and becomes negative, which obviously
corresponds to the elastic stress caused in the polymer ring by an
attempt to pull it away from the obstacle with which the ring is
entangled.

\subsection{Ring polymer entangled with a finite size disc}\label{sec:polymer_around_disc}

For the obstacle of finite radius $b$, we were unable to obtain
exact answers similar to Eqs. (\ref{eq:inversensquare}) or
(\ref{eq:rasstoyanie}).  All we can do for this case is to resort
to the asymptotic calculations.  One of the advantages of the
finite size obstacle model is that it allows to examine
\emph{both} asymptotics, we call them loose entanglement and tight
entanglement, respectively.  The former regime is realized when
the size of the obstacle $b$ is smaller than typical polymer coil
dimension $a t^{1/2}$, and, moreover, when minimal length
necessary to make $n$ turns, $2 \pi b n$, is still small compared
to $a t^{1/2}$: $n b \ll a t^{1/2}$.  In this case, calculations
are similar to those of Section \ref{sec:SpitzerDISC}.  In the
opposite extreme, when $n b \gg a t^{1/2}$, polymer has to be
significantly stretched out to make all $n$ turns.  This
corresponds to the far tail of winding angle distribution, which
is usually not examined and which we did not consider in Section
\ref{sec:SpitzerDISC}.

\subsubsection{Loose entanglement}

All we can do for this case is to resort to the asymptotic
calculations similar to those of Section \ref{sec:SpitzerDISC}. In
fact, the calculations are almost identical, and at the end they
return essentially the result (\ref{eq:dva_konca_fixed}), with the
only difference in the normalization factor. Specifically, the
probability to have linking number $n$ is proportional to
\be W_n \propto \frac{\pi x_n \sinh \pi x_n + 2 \left( 1 - \cosh
\pi x_n \right) }{\left( 1 - \cosh \pi x_n \right)^2} \ .\ee
where the omitted normalization factor must be defined such that
$\sum_{n=-\infty}^{\infty} W_n = 1$, and where
\be x_n = \frac{4 \pi n }{\ln \frac{t a^2 }{ b^2 }} \ . \ee
Similarly, although we cannot find the exact expression for the
value of $\sigma_n(t)$, but the estimate reads $ \sigma_n (t) \sim
\left[\pi b^2 + r^2  W_n \right]_{r \sim a \sqrt{t}}$.

Thus, quantities such as $W_n$ and $\sigma_n(t)$ decay
exponentially at very large $n$, and the characteristic $n$ where
exponential decay starts is about $\ln ( t a^2 / b^2)$.  This
latter quantity estimates also the characteristic linking number
in another sense, defined as the root mean squared, $\sqrt{
\langle n^2 \rangle}$.  This is an interesting and somewhat
unexpected result. Indeed, one could have naively expected that
the characteristic value of $n$ should be proportional to $t a^2 /
b^2$.  Indeed, we expect that one turn around the obstacle should
be similar to walking a distance about $2 \pi b \sim b$, which
requires a time $\tau \sim b^2 / a^2$, implying the number of
turns to be about $t / \tau$.  Instead, we are getting something
like $\ln (t / \tau )$. This happens because a large portion of
the chain length deviates much further away from the obstacle than
$b$, and it makes turns around a much larger circumference.  This
once again suggests that knot tightening \cite{Kardar_tightening}
occurs even in this case of a disc-like obstacle with excluded
volume.

Similar to our discussion in section \ref{sec:Spitzer_b_to_zero},
we can understand what happens when $b \to 0$.  In this case,
there appears a wide interval of polymer chain lengths $b^2/a^2
\ll t \ll a^2 / b^2$ in which, say, $\sigma_n(t)$ decays only as
$1/n^2$ in the wide interval of $n$, up to a large value of $n$ of
about $\ln (a / b)$.

\subsubsection{Tight entanglement}

%
%

\begin{figure}[ht]
\centerline{\scalebox{0.6} {\includegraphics{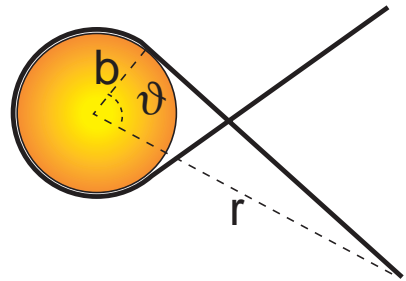}}}
\caption{Tight entanglement, or "ray optics" limit.  In this
figure, for the ease of drawing, we assume that the polymer makes
just a little more than one turn around the obstacle, while its
ends are fixed at the given points.  Distance to one end $r$ and
the corresponding angle $\vartheta$ are shown in the figure;
similar distance $r^{\prime}$ and angle $\vartheta^{\prime}$ are
not shown to simplify the figure.}\label{fig:tight}
\end{figure}

The tight entanglement regime is realized when $n^2 b^2 \gg t
a^2$.  In this case, polymer barely has enough length to make $n$
turns around the obstacle.  Obviously, the dominant polymer
conformations are those tightly wound around the obstacle.  This
regime is similar to ray optics \cite{Semenov}.  Indeed, if one
searches for the solution of diffusion equation
(\ref{eq:diffusion}) in the form $G =\exp \left[ s({\vec r}, t)
\right]$ and assumes that $s$ is (in a proper sense) a slowly
changing function, then the so-called eikonal equation for $s$
results:
\be -\frac{\partial s}{\partial t} = \frac{a^2}{6} \left(
\vec{\nabla} s \right)^2 \ . \label{eq:eikonal} \ee
For the system at hand, namely a polymer with winding angle
$\theta$ around the obstacle of radius $b$, this equation allows
for the exact solution:
\be s = \frac{3}{2 a^2} \frac{L^2}{t} \ , \label{eq:eikonsol1} \ee
where $L$ is the shortest distance between fixed ends consistent
with the topological constraint (that is, with the given winding
angle):
\be L = b (\theta - \vartheta - \vartheta^{\prime} ) + \sqrt{r^2 -
b^2} + \sqrt{\left. r^{\prime} \right. ^2 - b^2} \ .
\label{eq:eikonsol2} \ee
Here, $\vartheta$ and $\vartheta^{\prime}$ are determined by the
conditions $\cos \vartheta = b / r$ and $\cos \vartheta^{\prime} =
b / r^{\prime}$.  Both these conditions and the solution itself
are quite easy to establish based on the geometry presented in the
figure \ref{fig:tight}.  It is also easy to check by direct
differentiation that formulas
(\ref{eq:eikonsol1},\ref{eq:eikonsol2}) present an exact solution
of the eikonal equation (\ref{eq:eikonal}).

%
%
%
%

\subsection{Ring polymer inside a cavity}

Our discussion in the previous section is additionally illuminated
by the problem of a ring polymer entangled with an obstacle while
confined in a cavity of the radius $B$.  In this case, the result
(\ref{eq:otvet_cavity2}) directly applies, apart from the
replacement $\theta \to 2 \pi n$, and the proper normalization
factor. We see that in this case the characteristic value of $n$
is proportional to $ ta^2 / B^2$.  This number can be understood
by saying that polymer is confined to a tube of the width $D = B$
and length $L = n B$, where the typical $n$ must be determined
such that confinement entropy is similarly contributed by chain
squeezing across the tube and stretching along the tube (compare
similar arguments in \cite{knot_inflation}). The confinement
entropy is well known \cite{deGennes_book}, it is proportional to
$t a^2 / D^2 + L^2 / t a^2$, where the two terms correspond to the
two factors just mentioned - chain squeezing across the tube and
stretching along the tube, respectively.  Equating the two terms,
we arrive at $n \sim t a^2 / B^2$, as expected.

\section{Concluding remarks}\label{sec:conclusion}

We have focused on a special application, a toy
Edwards-Prager-Frisch model \cite{Edwards,Prager_Frisch}, and its
generalization.  It can be viewed as a model of an unsymmetrical
infinite catenane.  This catenane is formed from a random walk of
$t$ steps and ``entwined'' with infinite, rigid, closed structure
composed of two straight legs, which are separated at least by a
distance larger than $t a$, meeting at infinity.  As such it
allows one to extrapolate a limiting probability of catenation by
a closed random walk, which is consistent with earlier estimates.

This model is widely recognized as the simplest playground for
``statistical mechanics with topological constraints''
\cite{Edwards}. Unfortunately, no simple notable result had
previously come out of this model studies - except the very fact
that it is ``exactly solvable.'' We were lucky to find a couple of
such simple results. First, looking at the entanglement as an
element of ``annealed'' disorder, we found that the area
associated with all points around which closed random walk makes
exactly $n$ turns is equal to $\langle R^2 \rangle / 4 \pi n^2$,
where $\langle R^2 \rangle $ is the mean-square end-to-end
distance of the linear walk of the same length. Second, looking at
the entanglement as an element of ``quenched'' disorder, we found
that the mean-squared distance between an obstacle an an arbitrary
monomer of an $n$ times entangled ring is equal to $(1/6) \langle
R^2 \rangle \left[ 1 + 3 / 2 \pi n^2 \right]$.  Both results are
exact.  We have also found that the entanglement of a very long
polymer is very uneven, in the sense that it tends to segregate
into one very long loop, almost as long as the entire polymer, and
a number of much shorter loops.

The generalization of Edwards-Prager-Frisch model, in which an
obstacle is not a point, but a disc of finite radius $b$, as far
as we could tell, does not allow for an exact solution in any
useful closed form.  However, we were able to show that for
annealed loop the mean-squared linking number turns out to be
finite, of order $\langle n^2 \rangle \sim \ln (a/b)$.  As $b$
increases, the entanglement becomes increasingly tight, resulting
in all loops being of comparable length.

Our deliberations clearly reflect three aspects:  The first is the
fruitful nature of Spitzer's \cite{Spitzer} original insight for
both further theory and its applications.  The second is the value
of self-consistent approximations in the field when direct exact
calculations are precluded by mathematical difficulty.  And, last
but not least, the third is the usefulness of complete analysis of
those simplified models which do allow for mathematically exact
solutions.

\begin{acknowledgments}
The AG work was supported in part by the MRSEC Program of the NSF
under Award Number DMR-0212302.  Part of this work was performed
by AG while visiting Kavli Institute for Theoretical Physics,
University California - Santa Barbara;  this part was supported by
the National Science Foundation under Grant No. PHY99-07949.
Preprint of this work is published as NSF-KITP-03-044. HF
acknowledges the support under NSF grant DMR- 9628224.

\end{acknowledgments}

\appendix

\section{Weber integral}\label{sec_app:Weber_Integral}

First let us prove an auxiliary relation:
\be \int e^{- \alpha^2 \left({\vec r} - {\vec r}^{\prime}
\right)^2} \psi_{\kappa} ({\vec r}^{\prime}) d^2 {\vec r}^{\prime}
= \frac{\pi}{\alpha^2} e^{- \kappa / 4 \alpha^2} \psi_{\kappa}
({\vec r}) \ , \label{eq:auxiliary} \ee
where $\psi_{\kappa} ({\vec r})$ is an eigenfunction of the
Laplacian operator corresponding to the eigenvalue $-\kappa^2$:
$\Delta \psi = - \kappa^2 \psi ({\vec r})$.  To see this, we note
that $(\alpha^2 / \pi)  e^{- \alpha^2 \left({\vec r} - {\vec
r}^{\prime} \right)^2} $ is the Green function of the diffusion
equation in which $\alpha^2$ plays the role of $1 / D t$, with $D$
and $t$ being diffusion coefficient and time, respectively.
Therefore, this exponent can be written in terms of a bilinear
expansion
\be\frac{\alpha^2}{\pi} e^{- \alpha^2 \left({\vec r} - {\vec
r}^{\prime} \right)^2} = \sum_k \psi_k ({\vec r}) \psi_k ({\vec
r}^{\prime}) e^{- k^2 / \alpha^2 } \ , \ee
where the summation runs over the entire spectrum of the Laplacian
operator. By making a dot-product of both sides with
$\psi_{\kappa}$, we arrive at the result (\ref{eq:auxiliary}).

Let us now use the formula (\ref{eq:auxiliary}) choosing
$\psi_{\kappa} = J_{n} (\kappa r) e^{\imath n \theta}$.  At any
$n$, this is indeed one of the eigenfunctions of the Laplacian
operator, corresponding to the eigenvalue $-\kappa^2$.  Now, we
make use of the following relation
\be \int_0^{2 \pi} e^{ \alpha^2 \cos \phi + \imath \mu \phi} d
\phi = 2 \pi I_{\mu} (\alpha^2) \ , \ee
which is most frequently encountered as an integral representation
of the modified Bessel function $I_{\mu}$.
This relation leads to
\bea \int_0^{\infty} J_{\mu} (\kappa y) I_{\mu} (2 \alpha ^2 x y)
 & \times & e^{-\alpha^2 y^2} y \ \! dy = \nonumber \\ & = &
e^{\alpha^2 x^2 - \frac{\kappa^2}{4 \alpha^2} }
\frac{J_{\mu}(\kappa x)}{2 \alpha^2}  \ . \eea
Changing variables, we finally obtain the two formulations of the
Weber integral:
\bea \int_0^{\infty} J_{\mu} (\kappa y) I_{\mu} (\kappa^{\prime}
y) & \times & e^{-\alpha^2 y^2} y \ \! dy  = \nonumber \\ & = &
\frac{e^{\frac{\left. \kappa ^{\prime} \right.^2 - \kappa^2 }{ 4
\alpha^2} }}{2 \alpha^2}
 J_{\mu} \left( \frac{ \kappa \kappa^{\prime}}{2 \alpha^2} \right) \ , \nonumber \\
\int_0^{\infty} J_{\mu} (\kappa y) J_{\mu} (\kappa^{\prime} y) &
\times & e^{-\alpha^2 y^2} y \ \! dy = \nonumber \\ & = &
\frac{e^{-\frac{\left. \kappa ^{\prime} \right.^2 + \kappa^2 }{ 4
\alpha^2} }}{2 \alpha^2}
 I_{\mu} \left( \frac{ \kappa \kappa^{\prime}}{2 \alpha^2} \right) \ . \label{eq:nugnyi}  \eea
The latter formula is what needs to be used to go from Eq.
(\ref{eq:bilinear}) to Eq. (\ref{eq:afterWeber0}).

\section{Some properties of the functions $Z$}\label{appendix:Z}

\subsection{Orthogonality}

First of all, we want to prove here that the functions
$Z_{\mu}(\kappa r , \kappa b)$, as defined by the Eq.
(\ref{eq:Z}), are orthogonal and normalized:
\bea \int_b^{\infty} Z_{\mu} (\kappa r , \kappa b) Z_{\mu}
(\kappa^{\prime} r , \kappa^{\prime} b)  r d r & = &
\frac{1}{\kappa} \delta (\kappa - \kappa^{\prime})
\label{eq:answer2} \\  \int_0^{\infty} Z_{\mu} (\kappa r , \kappa
b) Z_{\mu} (\kappa r^{\prime} , \kappa b)  \kappa d\kappa & = &
\frac{1}{r} \delta (r - r^{\prime}) \label{eq:answer3} \eea

Note that when $b \to 0$, we have $Z_\mu (\kappa r, \kappa b)
\simeq J_{\mu} (\kappa r)$ (because in this case $Y_{\mu} (\kappa
b)$ is negative and large in absolute value), so in this limit
\textbf{both} equations (\ref{eq:answer2}) and (\ref{eq:answer3})
come back to the well known relation \cite{Morse_Feshbach}
\be \int_0^{\infty} J_{\mu} (\kappa r ) J_{\mu} (\kappa^{\prime} r
)  r d r  =  \frac{1}{\kappa} \delta (\kappa - \kappa^{\prime})
\label{eq:old_answer}  \ . \ee

In what follows, we derive the relations
(\ref{eq:answer2},\ref{eq:answer3}). For simplicity of notations,
it is easier to compute the normalization of the
functions
\be U_{\mu}(\kappa r , \kappa b) = -J_{\mu} (\kappa r) Y_{\mu}
(\kappa b) + J_{\mu} (\kappa b) Y_{\mu} (\kappa r) \ , \ee
from which the properties of $Z$ will follow automatically.
Derivation consists of two parts, one of them is trivial, and the
other is only slightly less trivial.

\subsubsection{Trivial part}

To begin with, there is a useful general formula
\be \int U_{\mu}^2 (\kappa r, \kappa b) \ \! r \ \! d r =
\frac{\kappa^2 r^2 - \mu^2}{2 \kappa^2} U_{\mu}^2 + \frac{r^2}{2
\kappa^2} \left( \frac{\partial U_{\mu}}{\partial r} \right)^2 \ ,
\label{eq:indeterminate} \ee
which is valid for any solution of Bessel equation, that is, for
any linear combination of $J_{\mu}(\kappa r)$ and $Y_{\mu}(\kappa
r) $, including the $U_{\mu}$.  The derivation of this formula can
be found in many places, for instance, \cite{Morse_Feshbach}.  One
way is to take $\left( r \partial U_{\mu} / \partial r \right)^2$
and differentiate it over $r$. Remembering that $U_{\mu}$
satisfies Bessel equation, it is easy to find
\be \frac{\partial}{\partial r} \left( r \frac{\partial
U_{\mu}}{\partial r} \right)^2 = 2 r^2 \left[ \frac{\mu^2}{r^2} -
\kappa^2 \right] U_{\mu} \frac{\partial U_{\mu}}{\partial r} \ ,
\ee
from which Eq. (\ref{eq:indeterminate}) follows automatically.

We cannot directly apply this formula for the case of an infinite
interval, because it yields the divergence (due to the $r^2
U_{\mu}^2$ term: $U_{\mu}$ decays only as $1 \sqrt{r}$ at large
$r$). Indeed, this is not surprising, since the answer
(\ref{eq:answer2}) contains a $\delta$-function.  Thus, what we
shall do is to consider first the finite width ring $b <r < B$,
with boundary condition $\left. U_{\mu} \right|_{r=B} = 0$.  In
the end, we shall send $B \to \infty$.

Assuming $\left. U_{\mu}(\kappa r, \kappa b) \right|_{r=B}$, we
get from (\ref{eq:indeterminate})
\be \int_b^B U_{\mu}^2 (\kappa r, \kappa b) r dr = \left[
\frac{r^2}{2 \kappa^2} \left( \frac{\partial U_{\mu}}{\partial r}
\right)^2 \right]_b^B \ . \ee
The derivative $\partial U_{\mu} / \partial r$ can be simplified,
because it is related to the Wronskian of $J_{\mu}(\kappa r)$ and
$Y_{\mu} (\kappa r)$, which is equal to $2/\pi \kappa r$.  Taking
into account the boundary condition
\be Y_{\mu} (\kappa b) J_{\mu} (\kappa B) = Y_{\mu}(\kappa B)
J_{\mu}(\kappa b) \ , \label{eq:boundary} \ee
we obtain
\bea \int_b^B U_{\mu}^2 (\kappa r, \kappa b) r dr & = &
\frac{2}{\kappa^2 \pi^2} \left[ \frac{J_{\mu}^2 (\kappa
b)}{J_{\mu}^2 (\kappa B)} - 1 \right] = \nonumber
\\ & = & \frac{2}{\kappa^2 \pi^2} \left[ \frac{Y_{\mu}^2
(\kappa b)}{Y_{\mu}^2 (\kappa B)} - 1 \right] \ . \label{eq:JY}
\eea

If we have two different values, $\kappa \neq \kappa^{\prime}$,
both satisfying boundary condition (\ref{eq:boundary}), then
\be \int_b^B U_{\mu} (\kappa r, \kappa b) U_{\mu} (\kappa^{\prime}
r, \kappa^{\prime} b) r dr  = 0  \ , \ee
as can be established either by proper integration by parts using
the Bessel equation, or by direct reference to the fact that these
$U$'s are the eigenfunctions of a Hermitian operator belonging to
different eigenvalues.  Thus, we can use the Kroneker symbol to
write
\bea && \int_b^B U_{\mu} (\kappa r, \kappa b) U_{\mu}
(\kappa^{\prime} r, \kappa^{\prime} b) r dr  = \nonumber \\ && \ \
\ \ \ \ \ \ \ \ \ \ = \frac{2}{\kappa^2 \pi^2} \left[
\frac{J_{\mu}^2 (\kappa b)}{J_{\mu}^2 (\kappa B)} - 1 \right]
\Delta_{\kappa \kappa^{\prime}}  \label{eq:finite} \ . \eea
Here, we sacrificed the beauty of symmetry and used the upper line
of the Eq. (\ref{eq:JY}); the same final answer is obtained from
the lower line.

\subsubsection{Slightly less trivial part}

We have to perform now the limit $B \to \infty$.  The difficulty
is that when $B$ changes, so does also $\kappa$, since it is
subject to boundary condition (\ref{eq:boundary}).  To circumvent
this problem, the following trick is suggested.  Let us choose
some particular value of $\kappa$, then boundary condition
(\ref{eq:boundary}) is satisfied by some discrete set of $B$
values.  Let us send $B \to \infty$ stepping over these specific
values and thus keeping $\kappa$ fixed.  Then, when $B$ is already
large enough, we can resort to the well known asymptotics
\bea J_{\mu} (x) & \simeq & \sqrt{\frac{2}{\pi x}} \cos \left[x -
\frac{\pi}{2} \left( \mu + \frac{1}{2} \right) \right] \ ,
\nonumber \\ Y_{\mu} (x) & \simeq & \sqrt{\frac{2}{\pi x}} \sin
\left[x - \frac{\pi}{2} \left( \mu + \frac{1}{2} \right) \right] \
. \label{eq:asymp} \eea
Then, formula (\ref{eq:finite}) yields
\bea && \int_b^B U_{\mu} (\kappa r, \kappa b) U_{\mu}
(\kappa^{\prime} r, \kappa^{\prime} b) r dr  =
\label{eq:finite1}\\ && \ \ \ = \frac{2}{\kappa^2 \pi^2} \left[
\frac{\pi \kappa B}{2} \frac{J_{\mu}^2 (\kappa b)}{\cos ^2
\left[\kappa B - \frac{\pi}{2} \left( \mu + \frac{1}{2} \right)
\right]} - 1 \right] \Delta_{\kappa \kappa^{\prime}} \nonumber  \
. \eea
On the other hand, we can also use asymptotics (\ref{eq:asymp})
for $J_{\mu}(\kappa B)$ and for $Y_{\mu}(\kappa B)$ to simplify
the boundary condition (\ref{eq:boundary});  by some easy
manipulations, we can re-write this boundary condition in the form
\be 1 + \left[ \frac{Y_{\mu}(\kappa b)}{J_{\mu}(\kappa b)}
\right]^2 = \frac{1}{\cos^2 \left[\kappa B - \frac{\pi}{2} \left(
\mu + \frac{1}{2} \right) \right]} \ .
\label{eq:simplified_boundary} \ee
Then, formula (\ref{eq:finite1}) yields to the leading order in
$B$:
\bea && \int_b^B U_{\mu} (\kappa r, \kappa b) U_{\mu}
(\kappa^{\prime} r, \kappa^{\prime} b) r dr  = \nonumber \\ && \ \
\ \ \ \ \ \ \ \ \ \ =  \frac{B}{\kappa \pi} \left[ J_{\mu}^2
(\kappa b) + Y_{\mu}^2 (\kappa b) \right] \Delta_{\kappa
\kappa^{\prime}} \label{eq:finite2} \eea

Finally, we argue that at large $B$ the Kronecker $\Delta$ should
be replaced with Dirac $\delta$ according to
\be \Delta_{\kappa \kappa^{\prime}} \to \frac{\pi}{B}
\delta(\kappa - \kappa^{\prime})  \ . \label{eq:Dirac} \ee
To make this conclusion, we switch to the view point in which $B$
can be arbitrary, while $-\kappa^2$ and $-\left. \kappa^{\prime}
\right.^2$ are the eigenvalues which depend on $B$.  Then, when $B
\to \infty$, the eigenvalues come closer to one another, with the
interval between neighboring $\kappa$ equal to $\pi / B$, as it is
clear from the asymptotics of Bessel functions (\ref{eq:asymp}).
Therefore, any sum involving Kroneker $\Delta$ can be transformed
into the integral
\be \sum_{\kappa^{\prime}} \ldots \Delta_{\kappa \kappa^{\prime}}
\to \int \ldots \Delta_{\kappa \kappa^{\prime}} \frac{d
\kappa^{\prime}}{\pi / B} \ , \ee
which means precisely (\ref{eq:Dirac}).

Taken together, equations (\ref{eq:finite2}) and (\ref{eq:Dirac})
yield the answer
\be \int_b^{\infty} U_{\mu} (\kappa r , \kappa b) U_{\mu}
(\kappa^{\prime} r , \kappa^{\prime} b)  r d r  = \frac{J_{\mu}^2
(\kappa b) + Y_{\mu}^2 (\kappa b)}{\kappa} \delta (\kappa -
\kappa^{\prime}) \label{eq:answer} \ee
which is essentially formula (\ref{eq:answer2}). Formula
(\ref{eq:answer3}) follows automatically from (\ref{eq:answer2})
and the the fact that functions $Z_{\mu}$ form a complete set,
which, in turn, follows from the very general spectral
consideration.

\subsection{Asymptotics of $Z$}

Here, we first briefly describe the derivation of the small
$\kappa$ asymptotics of $Z_{\mu}(\kappa r, \kappa b)$, Eq.
(\ref{eq:expandZ}).  Knowing that \cite{Morse_Feshbach}
\bea J_{\mu} (\xi) & \simeq & \frac{(\xi / 2)^{\mu}}{\Gamma ( 1+
\mu )} \ , \nonumber \\ Y_{\mu} (\xi) & \simeq & \frac{(\xi /
2)^{\mu}}{\Gamma ( 1+ \mu )}\cot \pi \mu - \frac{(2 / \xi
)^{\mu}}{\Gamma ( 1- \mu )} \frac{1}{\sin \pi \mu}
\label{eq:small_xi} \eea
at $\xi^2 \ll 1 + \mu$, and using the identity
\be \Gamma ( 1 + \mu) \Gamma (1 - \mu ) \sin \pi \mu = \pi \mu \ ,
\ee
and not resorting to any further approximations, we arrive at the
Eq. (\ref{eq:expandZ}).

For completeness, we also mention the large $\kappa$ asymptotics
of $Z$, which turn out to be particularly nice:
\be Z_{\mu}(\kappa r, \kappa b) \simeq \sqrt{\frac{2}{\pi \kappa r
}} \sin \left( \kappa (b - r) \right) \ . \ee

\section{Alternative representation of the Green's function}\label{appendix:LaplaceFourier}

Consider again the diffusion equation, which at $t > t^{\prime}$
reads
\be \frac{\partial G}{\partial t} = \frac{a^2}{4} \Delta G \ , \ee
subject to the initial and the boundary conditions
\be \left. G \right|_{t=0} = \delta (\vec{r} - \vec{r}^{\prime} )
\ , \ \ \ \left. G \right|_{r=b} = 0 \ , \ \ \ \left. G \right|_{r
\to \infty} \to 0 \ . \ee
%
Let us denote
%
%
%
%
%
\be G_{p,\mu}(r,r^{\prime}) = \int_0^{\infty} e^{-pt}
\int_{-\infty}^{+\infty} e^{\imath \mu \theta} G(t,
\vec{r},\vec{r}^{\prime}) d \theta d t \ . \ee
%
%
%
This satisfies
\bea \frac{1}{r^2} \frac{\partial}{\partial r} r^2
\frac{\partial}{\partial r}G_{p,\mu}(r,r^{\prime}) &-& \left[
\frac{4 p}{a^2} + \frac{\mu^2}{r^2} \right]
G_{p,\mu}(r,r^{\prime}) = \nonumber \\ & = & - \frac{4}{r a^2}
\delta(r -r^{\prime})  \ . \label{eq:diff_Laplace} \eea
The solution of this equation is the linear combinations of the
Bessel functions $K_{\mu}$ and $I_{\mu}$.  Making it to satisfy
the boundary conditions at $r=0$, $r \to \infty$, and at $r =
r^{\prime}$ (the latter dictated by the $\delta$-function), one
arrives at
\begin{widetext}
\be G_{p,\mu}(r,r^{\prime}) = \left\{\begin{array}{lcr}
\frac{4}{a^2} \frac{K_{\mu} \left( \frac{2  r^{\prime}
\sqrt{p}}{a} \right)}{K_{\mu} \left( \frac{2  b \sqrt{p}}{a}
\right)}  \left[I_{\mu} \left( \frac{2  r \sqrt{p}}{a}
\right)K_{\mu} \left( \frac{2  b \sqrt{p}}{a} \right) - I_{\mu}
\left( \frac{2 b \sqrt{p}}{a} \right)K_{\mu} \left( \frac{2 r
\sqrt{p}}{a} \right) \right] &{\rm when} & r < r^{\prime}\\ \\
\frac{4}{a^2} \frac{K_{\mu} \left( \frac{2  r \sqrt{p}}{a}
\right)}{K_{\mu} \left( \frac{2  b \sqrt{p}}{a} \right)}
\left[I_{\mu} \left( \frac{2 r^{\prime} \sqrt{p}}{a}
\right)K_{\mu} \left( \frac{2  b \sqrt{p}}{a} \right) - I_{\mu}
\left( \frac{2 b \sqrt{p}}{a} \right)K_{\mu} \left( \frac{2
r^{\prime} \sqrt{p}}{a} \right) \right] & {\rm when} & r >
r^{\prime}
\end{array} \right. \ . \label{eq:Laplace_otvet} \ee
\end{widetext}
As expected, this is the symmetric function of $r$ and
$r^{\prime}$.  This formula was already obtained in
\cite{RudnickHu1} (equation (2.9) of that work).

\begin{figure}
\centerline{\scalebox{0.6} {\includegraphics{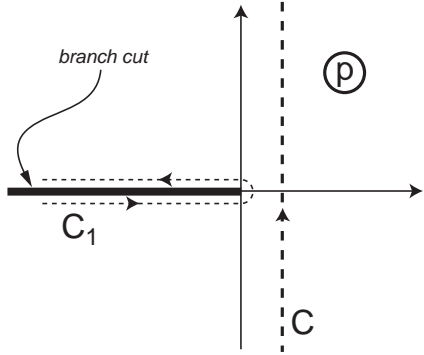}}}
\caption{Integration contours on the complex $p$-plane.
Explanations are in the text. }\label{fig:contour}
\end{figure}

What we should do now is to invert the respective Laplace and
Fourier transforms:
\be G(r,r^{\prime},\theta,t) = \frac{1}{(2 \pi)^2 \imath}
\int_{-\infty}^{\infty}  \int_{\cal C} e^{-\imath \mu \theta} e^{p
t} G_{p,\mu}(r,r^{\prime}) d p d \mu \ , \ee
where ${\cal C}$ is the vertical contour in the plane of complex
variable $p$ which should be to the right of all singularities of
$G_{p,\mu}$.  Knowing the explicit expression of $G_{p,\mu}$, Eq.
(\ref{eq:Laplace_otvet}), we see that it has the singularity at
$p=0$.  This singularity is due to both the branch point of
$\sqrt{p}$ and the singular behavior of many Bessel functions at
zero.  Then, it is convenient to place the branch cut along the
negative real axis in complex $p$-plane, and then to deform the
contour from ${\cal C}$ to ${\cal C}_1$, as shown in the Figure
\ref{fig:contour}.  Then, because of the branch cut, on the lower
side of the contour ${\cal C}_1$ we have $p= e^{-\imath \pi}
\left| p \right|$, while on the upper side we have $p = e^{\imath
\pi} \left| p \right|$.  Furthermore, instead of $\left| p
\right|$, it is convenient to introduce the new variable,
$\kappa$, such that $\left| p \right| = \kappa^2 a^2 /4$.  Then,
integrals along the lower and along the upper sides of the contour
${\cal C}_1$ are each represented by integration from $0$ to
$\infty$ over $\kappa$.  We can combine these two integrals
together, and then simple algebra yields
\begin{widetext} \bea G(r,r^{\prime},\theta,t) & = & \frac{2 \imath}{(2 \pi)^2
} \int_{-\infty}^{\infty}  \int_{0}^{\infty} e^{-\imath \mu
\theta} e^{-\kappa^2 a^2 t/4} \left\{ \frac{K_{\mu} \left( \imath
\kappa r^{\prime}  \right)}{K_{\mu} \left( \imath \kappa  b
\right)} \left[I_{\mu} \left( \imath \kappa  r \right)K_{\mu}
\left( \imath \kappa  b  \right) - I_{\mu} \left( \imath \kappa b
\right) K_{\mu} \left( \imath \kappa r \right) \right] \right. -
\nonumber
\\ & - & \left. \frac{K_{\mu} \left( - \imath \kappa r^{\prime}
\right)}{K_{\mu} \left(- \imath \kappa  b \right)} \left[I_{\mu}
\left( - \imath \kappa  r \right)K_{\mu} \left( - \imath \kappa  b
\right) - I_{\mu} \left(- \imath \kappa b \right) K_{\mu} \left( -
\imath \kappa r \right) \right] \right\} \kappa d \kappa d \mu \ .
\label{eq:Laplace_otvet1} \eea
%
%
\end{widetext}

The expression in curly brackets here can be simplified using the
following three relations:
\bea && I_{\mu} \left( \imath \kappa  r \right)K_{\mu} \left(
\imath \kappa  b  \right) - I_{\mu} \left( \imath \kappa b \right)
K_{\mu} \left( \imath \kappa r \right) = \nonumber \\ & = &
I_{\mu} \left( - \imath \kappa  r \right)K_{\mu} \left( - \imath
\kappa  b \right) - I_{\mu} \left(- \imath \kappa b \right)
K_{\mu} \left( - \imath \kappa r \right) = \nonumber \\ & = & -
\frac{\pi}{2} \left[ J_{\mu} \left( \kappa r \right)Y_{\mu} \left(
\kappa b \right) - J_{\mu} \left(
 \kappa b \right) Y_{\mu} \left( \kappa r \right) \right] \ ,\eea
and
\be K_{\mu} \left( \imath \kappa b \right) K_{\mu} \left( - \imath
\kappa  b \right) = \left( \frac{\pi}{2} \right)^2 \left[
J_{\mu}^2 \left( \kappa b \right) + Y_{\mu}^2 \left( \kappa b
\right) \right] \ , \ee
and
\bea && K_{\mu} \left( \imath \kappa r^{\prime}  \right) K_{\mu}
\left( - \imath \kappa  b \right) - K_{\mu} \left( - \imath \kappa
r^{\prime} \right) K_{\mu} \left( \imath \kappa  b \right) =
\nonumber \\ &=& - \frac{\pi^2 \imath}{2} \left[ J_{\mu}\left(
\kappa r^{\prime} \right) Y_{\mu}\left( \kappa b \right) -
J_{\mu}\left( \kappa b \right) Y_{\mu}\left( \kappa r^{\prime}
\right) \right] \ . \eea
Using these three results, we directly see that the formula
(\ref{eq:Laplace_otvet1}) gets transformed into
(\ref{eq:bilinearZ}).  This can, of course, be considered as
another proof of normalization conditions for $Z$-functions, Eq.
(\ref{eq:answer2},\ref{eq:answer3}).

\section{Integration over the coordinate $r^{\prime}$ of the trajectory end}\label{sec_app:integrate_over_end}

Most easily, integration in (\ref{eq:integrate_over_end}) can be
addressed using Eq. (\ref{eq:afterWeber0}).  Indeed, whatever is
the value of $r$, we should consider the limit of large $t$ in the
sense that $t a^2 \gg r^2$; that means, by the time $t$ the walker
should have traveled from its starting point typically much
further than to the origin, ${\cal O}$.   Then, we note that
although the integration over $r^{\prime}$ runs to infinity, the
integral is dominated by $r^{\prime}$ up to about $a \sqrt{t}$,
because of the truncation by the exponential factor $e^{- \left.
\left. r^{\prime} \right. ^2 \right/ a^2 t}$.  Accordingly, the
argument $\xi = 2 r r^{\prime} / a^2 t$ of $I_{\mu}$ in the Eq.
(\ref{eq:afterWeber0}) is small, and we can use the expansion
 $I_{\mu} (\xi) \simeq \left. \left( \xi / 2 \right)^{\mu} \right/ \Gamma ( 1 + \mu ) $.  Upon
integration over $r^{\prime}$, this yields
\bea W(\theta) & \propto & e^{- r^2 / a^2 t} \times \nonumber
\\ & \times & \int_0^{\infty} \cos ( \mu \theta) \frac{\Gamma \left( 1 +
\frac{\mu }{2} \right)}{\Gamma ( 1+ \mu)} \left( \frac{ r^2 }{ a^2
t} \right)^{\frac{\mu}{2}} d \mu \ .
\label{eq:mu_is_small_when_integr_over_end} \eea
Now, we have to remember that $ r^2 / a^2 t \ll 1$, which means
that the latter integral is dominated by small $\mu$, more
specifically by $\mu$ up to about $1/ \left| \ln \left(  r^2 /
a^2 t \right) \right|$.   Replacing both $\Gamma$-functions with
unity leads then to
\bea W(\theta) & \propto & \frac{\frac{1}{2} \ln \left( \frac{ a^2
t}{ r^2 } \right)}{\left( \frac{1}{2} \ln \left( \frac{ a^2 t}{
r^2 } \right) \right)^2 + \theta ^2}  \simeq \nonumber \\ & \simeq
& \frac{\frac{1}{2} \ln t }{\left( \frac{1}{2} \ln t \right)^2 +
\theta^2} \ , \eea where the latter transformation is justified
again because $t$ is large.  Thus, the resulting distribution is
indeed independent of $r$, and it is nothing else but the Spitzer
law, Eq. (\ref{eq:Spitzer}).

\section{Proof that $W$ is the
probability}\label{sec_app:normalization_of_W}

Here, we check that $W$ satisfies the normalization condition as
the probability:
\be \sum_{n=-\infty}^{\infty} W \left(\theta + 2 \pi n, \frac{3 r
r^{\prime}}{a^2 t } \right) = 1 \ , \label{eq:normalization} \ee
which also means that identification of the points $\theta$ and
$\theta + 2 \pi n$ erases all the topological information.

First, let us denote for brevity $z= 2 r r^{\prime} / a^2 t $, and
then we write
\bea && \sum_{n=-\infty}^{\infty} W \left(\theta + 2 \pi n, z
\right) = \nonumber \\ & = & 2 e^{-z \cos \theta} \int_0^{\infty}
\left[\sum_{n=-\infty}^{\infty} \cos \left( \left( \theta + 2 \pi
n \right) \mu \right) \right] I_{\mu} (z) d \mu = \nonumber \\ & =
& e^{-z \cos \theta} \int_{-\infty}^{\infty}
\left[\sum_{n=-\infty}^{\infty} \cos \left( \left( \theta + 2 \pi
n \right) \mu \right) \right] I_{\left| \mu \right|} (z) d \mu \ .
\eea
Here, integration is expanded over all $\mu$, both positive
and negative, the price being the absolute value of $\mu$ serving
as an index of $I_{\left| \mu \right|}$.  The expression in the
square brackets can be easily transformed using the identity
\be \sum_{k=-\infty}^{\infty} e^{2 \pi \imath k t} =
\sum_{m=-\infty}^{\infty} \delta (t-m) \ee
%
Thus, we write
\bea && \sum_{n=-\infty}^{\infty} \cos \left( \left( \theta + 2
\pi n \right) \mu \right)  =  \nonumber \\ & = & \frac{1}{2}
\sum_{n=-\infty}^{\infty} \left[ e^{\imath \mu (\theta + 2 \pi n)}
+ e^{\imath \mu (- \theta - 2 \pi n)} \right]  =  \nonumber \\ & =
& \frac{1}{2} \sum_{n=-\infty}^{\infty} \left[ e^{\imath \mu
(\theta + 2 \pi n)} + e^{\imath \mu (- \theta + 2 \pi n)} \right]
 =  \nonumber \\ & = & \frac{1}{2} \sum_{n=-\infty}^{\infty} e^{2
\pi \imath n \mu} \left[ e^{\imath \mu \theta} + e^{- \imath \mu
\theta } \right] =  \nonumber \\ & = & \cos ( \mu \theta)
\sum_{n=-\infty}^{\infty} e^{2 \pi \imath n \mu} =  \nonumber
\\ & = & \cos ( \mu \theta) \sum_{m=-\infty}^{\infty} \delta (\mu - m)
\ . \eea
We also use the two identities \cite{GR}:
\be I_{\nu} = e^{-\imath \nu \pi /2} J_{\nu} (\imath z) \ee
(formula 8.406.3 in \cite{GR}) and
\be e^{\imath z \cos \phi} = J_0 (z) + 2 \sum_{k=1}^{\infty}
\imath ^k J_k (z) \cos (k \phi) \ee
(formula 8.511.4 in \cite{GR}). This yields
\bea && \sum_{n=-\infty}^{\infty} W \left(\theta + 2 \pi n, z
\right) = \nonumber \\ & = &  e^{-z \cos \theta}
\sum_{m=-\infty}^{\infty} e^{\imath m \theta }  I_{\left| m
\right|} (z)  = \nonumber \\ & = &  e^{-z \cos \theta}
\sum_{m=-\infty}^{\infty} e^{\imath m \theta - \imath \left| m
\right| \pi /2} J_{\left| m \right|} ( \imath z)  = \nonumber \\ &
= &  \left[ \underbrace{ J_0 (\imath z) + 2 \sum_{m=1}^{\infty}
\cos (m \theta - m \pi ) \imath^m J_m (\imath z) }_{e^{\imath
(\imath z) \cos (\theta - \pi) }}\right] \times  \nonumber \\ &
\times & e^{- z \cos \theta} =  1 \ . \eea
This completes the proof.

\end{document}